\documentclass[aps,prl,reprint,noshowpacs,noshowkeys,floatfix%
]{revtex4-2}

\usepackage{graphicx}

\usepackage{amsmath,amssymb}

\usepackage[backref=none]{hyperref}
\usepackage[usenames,dvipsnames]{color}
\definecolor{MyDarkBlue}{rgb}{0,0.1,0.7}
\hypersetup{pdfborder={0 0 0},colorlinks,breaklinks=true,
	urlcolor={MyDarkBlue},citecolor={MyDarkBlue},linkcolor={MyDarkBlue}
}

\newcommand{\heff}{\hbar_{\mathrm{eff}}}

\newcommand{\hamiltonian}{\hat{\mathcal{H}}}

\newcommand{\cre}[1]{\hat{a}^\dagger_{#1}}
\newcommand{\ann}[1]{\hat{a}_{#1}}

\newcommand{\nt}{\tilde{N}}

\newcommand{\eh}[1]{e^{#1}}

\newcommand{\bplus}{\hat{b}_{+}}
\newcommand{\bminus}{\hat{b}_{-}}

\begin{document}
	\title{Emergence of a Renormalized $1/N$ Expansion in Quenched Critical Many-Body Systems}
	\newcommand{\RegensburgUniversity}{Institut f\"ur Theoretische Physik, 
		Universit\"at Regensburg, D-93040 Regensburg, Germany}
	\author{Benjamin Geiger}
	\affiliation{\RegensburgUniversity}
	\author{Juan Diego Urbina}
	\affiliation{\RegensburgUniversity}
	\author{Klaus Richter}
	\affiliation{\RegensburgUniversity}
	\date{March 25, 2021}
	
	\begin{abstract} 
		We consider the fate of $1/N$ expansions in unstable many-body quantum systems, as realized by a quench across criticality, and show the emergence of ${\rm e}^{2\lambda t}/N$ as a renormalized parameter ruling the quantum-classical transition and accounting nonperturbatively for the local divergence rate $\lambda$ of mean-field solutions.
		In terms of ${\rm e}^{2\lambda t}/N$, quasiclassical expansions of  paradigmatic examples of criticality, like the self-trapping transition in an integrable Bose-Hubbard dimer and the generic instability of attractive bosonic systems toward soliton formation, are pushed to arbitrarily high orders.
		The agreement with numerical simulations supports the general nature of our results in the appropriately combined long-time $\lambda t\to \infty$ quasiclassical $N\to \infty$ regime, out of reach of expansions in the bare parameter $1/N$.
		For scrambling in many-body hyperbolic systems, our results provide formal grounds to a conjectured multiexponential form of out-of-time-ordered correlators.
	\end{abstract}
	
	
	\maketitle
	
	The semiclassical (small $\hbar$) expansion provides a tool to address in a systematic way quantum effects on observables admitting a power expansion around their classical values where $\hbar=0$ \cite{Wigner1932,Glauber2007}. 
	It comes as no surprise then that such expansions have achieved a prominent role in the study of both the quantum-classical transition itself \cite{Zurek2003} and the physics of systems in the mesoscopic regime \cite{Richter2000,Brack2003}. 
	With the impressive advances in the coherent preparation, control, and manipulation of ever larger quantum systems, this mesoscopic regime where the system remains coherent but where typical classical actions are large compared with $\hbar$ keeps growing and, with it, the range of applications of these methods \cite{Hornberger2012}. 
	The extensive use of phase space methods based on Wigner-Moyal calculus in its several variants, like high-temperature expansions in statistical mechanics \cite{Wigner1932,Polkovnikov2010,Rundle2019}, truncated Wigner approximation describing the dynamics of cold atoms \cite{Sinatra2002,Norrie2006,Deuar2007,Blakie2008}, Weyl-Kirkwood expansions in nuclear physics \cite{Brack2003,Fujiwara1982}, and others, show the breadth and power of quasiclassical expansions \footnote{Quasiclassical refers to neglecting interfering classical solutions.} even in the presence of decoherence \cite{Gasenzer2005,Witthaut2011,Kordas2013,Ivanov2013}.
	
	When lifted into the realm of interacting many-body systems admitting a well-defined mean-field (MF) description in terms of bosonic order parameters, the quasiclassical expansion is formally constructed by means of the key identification of the number of particles $N$ as inverse effective Planck constant, namely $\heff=1/N$, where quantum fluctuations around the MF limit assume the form of expansions in powers of $1/N$ \cite{Chatterjee1990,Zibold2010,Polkovnikov2011,Pappalardi2018}.
	In particular, matrix elements of time-dependent operators in the Heisenberg picture are expected to have a (at least asymptotic) $1/N$ expansion. 
	The limitations of such a bare large-$N$ expansion become evident when the dynamics of the observables is driven by a quench across a phase transition \cite{Mitra2018}, here defined as an instability of the MF when changing a control parameter \cite{Emary2003,Caprio2008,Bastidas2014,Stransky2014,Bastarrachea-Magnani2016,Rubeni2017,Pappalardi2018,Hummel2019b}.
	The interplay between local MF instability, measured by an imaginary Bogoliubov frequency $\lambda$, and quantum fluctuations makes the Wigner-Moyal expansion valid only before the onset of nonperturbative quantum interference effects  \cite{Gutzwiller1990} at a timescale $t_{E} \sim \log N/(2\lambda)$ \cite{Chirikov1981,Rammensee2018} parametrically small in $N$.
	In practice, however, the quasiclassical expansion (and its characteristic Taylor-like form in $t$ and $1/N$) breaks down well before this fundamental limitation due to the uncontrolled complexity of high-order Moyal expansions.
	\begin{figure}[ttt]
		\includegraphics[width=\linewidth]{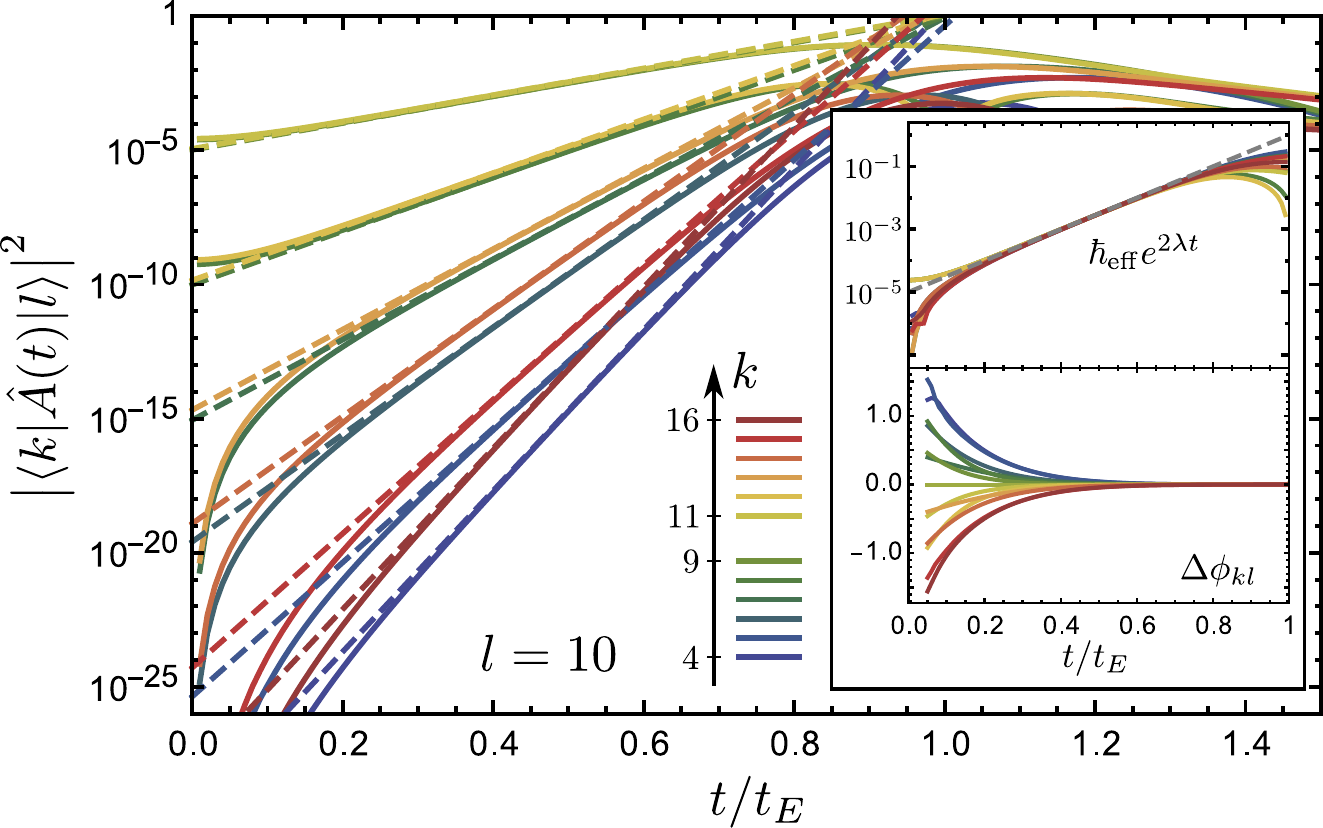}
		\caption{%
			Exponential growth of matrix elements, Eq.~\eqref{eq:matrix_elements}, with $\hat{A}=\hat{z}+\hat{z}^2$ after an interaction quench from $\alpha=0$ to $\alpha=2.5$ for $N=10^5$ particles in the Bose-Hubbard dimer [Eqs.~\eqref{eq:BH_dimer} and \eqref{eq:Hmf}, $\hat{z}=(\hat{n}_1-\hat{n}_2)/2N$]. The main panel shows the absolute squares for $l=10$ and $k\neq l$ ranging from $4$ to $16$.
			Solid lines represent the numerical data, while the dashed lines show the predicted dominant scaling, Eq.~\eqref{eq:matrix_elements}.
			Inset: shows a collapse of the absolute values (top) and the phases (bottom) using the prediction for the coefficients $c_{kl}$ \cite{SM}.
		}
		\label{fig:two-site_BH_comparison}
	\end{figure}
	\nocite{Hummel2019b,Jasiulewicz2003,James1962,Fetter1971}
	
	In this Letter, we address this early breakdown of $1/N$ expansions around MF instabilities and show that it can be pushed up to $t\sim t_{E}$ by introducing a renormalized small parameter $\eh{\lambda t}/\sqrt{N}$ that subsumes the effect of local hyperbolicity.
	This is demonstrated by suppressing leading orders in the $1/N$ expansion of Heisenberg operators using (approximate) harmonic oscillator states $|k\rangle$ localized at the instability, leading to the scaling
	\begin{equation}
	\langle k|\hat{A}(t)|l\rangle \sim c_{kl}\left(e^{\lambda t}/\sqrt{N}\right)^{|k-l|}=c_{kl}\eh{\lambda(t-t_E)|k-l|},
	\label{eq:matrix_elements}
	\end{equation}
	in the regime $\lambda t \gg 1, N \to \infty$, with $N$-independent constants $c_{kl}$. 
	Beyond the shortest timescale $1/\lambda$ (often associated with single-particle scattering or relaxation) matrix elements exponentially increase, shifted by $t_E$.
	This general prediction provides the precise way in which $t$ and $N$ commonly determine this increase, effectively absorbing the effect of local MF instability within the $1/N$ expansion in a single, renormalized expansion parameter.
	Equation~\eqref{eq:matrix_elements} is compared against extensive numerical simulations for the paradigmatic case of the self-trapping transition in the integrable Bose-Hubbard dimer \cite{Julia-Diaz2010,Zibold2010,Rautenberg2019,Kunkel2019} in Fig.~\ref{fig:two-site_BH_comparison}, which extends to spin-$1/2$ chains with long-range interactions in the symmetric subspace \cite{Pappalardi2018}.
	Later, we will report similar results for the nonintegrable three-site model \cite{Nemoto2000,Franzosi2003,Mossmann2006,Oelkers2007} with close relation to the generic instability of attractive Bose gases toward soliton formation \cite{Kanamoto2005,Khaykovich2002}, while similar considerations apply to any system with a well-defined classical limit, including chains with high ($>3/2$) local spin \cite{Akila2017}.
	
	Our analysis is based on the generic behavior of time-dependent perturbative expansions of interacting bosonic systems applied to a quench around criticality.
	It is illustrated for definiteness using the two-site Bose-Hubbard Hamiltonian
	\begin{equation}
	\hat{H}=-J\left[\cre{1}\ann{2}+\cre{2}\ann{1}\right]+\frac{U}{2}\left[(\cre{1})^2(\ann{1})^2+(\cre{2})^2(\ann{2})^2\right].
	\label{eq:BH_dimer}
	\end{equation}
	It allows for a detailed analysis due to its low level of complexity, while generalization to an arbitrary number of sites can be found in the Supplemental Material \cite{SM}.
	The steps of the calculation are the same in all cases:
	First, a MF model has to be identified as a formal classical limit, using the (conserved) inverse particle number $1/N$ as an effective Planck constant.
	Then, an expansion of the MF Hamiltonian around the prequench global minimum is used to obtain a well-controlled low-energy description of the system via canonical quantization.
	An expansion of the postquench unstable dynamics around the very same point is then the starting point for a perturbative analysis that finally reveals the renormalized expansion parameter, leading to the scaling in Eq.~\eqref{eq:matrix_elements}.

	\textit{Quantization of the mean field. --- }
	As $N$ is conserved, the MF dynamics is described using the occupation imbalance and the conjugate phase only, leading to the energy per particle \cite{SM}
	\begin{equation}
	\frac{H_\mathrm{MF}}{\nt J}=h(z,\varphi)-1-\alpha\left(\frac{1}{2}+\frac{1}{\nt}-\frac{1}{\nt^2}\right),
	\label{eq:Hmf}
	\end{equation}
	with the classical MF dynamics only determined by the Josephson Hamiltonian
	\begin{align}
	h(z,\varphi)=1-\sqrt{1-4z^2}\cos\varphi-2\alpha z^2.
	\label{eq:h}
	\end{align}
	In Eqs.~\eqref{eq:Hmf} and \eqref{eq:h}, $z=(n_1-n_2)/(2\nt)$ is the relative occupation imbalance, with $\nt=N+1\approx N$, $\varphi$ is the conjugate phase,
	and the interaction is scaled as $\alpha=-U\nt/2J$ with $\alpha>0$ for attractive coupling. The global energy minimum at $z=\varphi=0$ for couplings $\alpha\leq 1$ is replaced by a hyperbolic fixed point for $\alpha>1$.
	
	We first consider the quadratic	approximation of Eq.~\eqref{eq:h} around the global minimum for $\alpha<1$, described by a harmonic oscillator
	\begin{equation}
	h_0(z,\varphi)=\frac{\varphi^2}{2}+\frac{\omega^2z^2}{2}, {\rm \ \ }\omega=2\sqrt{1-\alpha}.
	\label{eq:h0}
	\end{equation}
	The states $|k\rangle$ in Eq.~\eqref{eq:matrix_elements} are thus harmonic oscillator states. For $\alpha>1$ we use the full model
	\begin{equation}
	h(z,\varphi)=\frac{\varphi^2}{2}-\frac{\lambda^2z^2}{2}+v(z,\varphi),
	\label{eq:h_expansion}
	\end{equation}
	with $v$ at least cubic in $(z,\varphi)$ and
	\begin{equation}
	\lambda=2\sqrt{\alpha-1}
	\end{equation}
	being the instability of the hyperbolic fixed point.
	
	The MF Hamiltonians \eqref{eq:h0} and \eqref{eq:h_expansion} are then quantized by replacing the variables $z,\varphi$ by operators (using symmetric ordering) and requiring the commutator relation
	\begin{equation}
	[\hat{z},\hat{\varphi}]=i\heff=\frac{i}{\nt}.
	\end{equation}
	
	\textit{Interaction picture. --- }
	Using
	\begin{equation}
	\hat{A}(t)=e^{\frac{i}{\heff} (\hat{h}-\hat{v})}\hat{A}e^{-\frac{i}{\heff} (\hat{h}-\hat{v})},
	\end{equation}
	we treat the postquench dynamics generated by the quadratic part of $\hat{h}=h(\hat{z},\hat{\varphi})$ exactly.
	A Heisenberg operator (denoted with subindex $H$) can then be formally expanded as \cite{Franson2002}
	\begin{align}
	\hat{A}_H(t)
	=&
	\sum_{n=0}^{\infty}\frac{1}{(i\heff)^n}\int_{0}^{t}\!dt_1\int_{0}^{t_{1}}\!dt_2\,\dots\int_{0}^{t_{n-1}}\!dt_n\,
	\nonumber
	\\ &\quad
	\big[\big[\dots\big[\big[\hat{A}(t),\hat{v}(t_1)\big],\hat{v}(t_2)\big],\dots\big],\hat{v}(t_n)\big],
	\label{eq:Heisenberg_operator_expansion}
	\end{align}
	where the zero-order term is defined as $\hat{A}(t)$.
	Equation \eqref{eq:Heisenberg_operator_expansion} can be used as a perturbation expansion when
	both $\hat{z}$ and $\hat{\varphi}$ can be considered as small, as is the case for the prequench eigenstates that have a characteristic extent $\sim \sqrt{\heff}$ in both $z$ and $\varphi$.
	We make this explicit by defining the Hermitian operators
	\begin{equation}
	\hat{b}_\pm=\frac{1}{\sqrt{2\lambda\heff}}\left(\lambda \hat{z}\pm \hat{\varphi}\right),
	\qquad
	\big[\hat{b}_-,\hat{b}_+\big]=i,
	\end{equation}
	with a trivial time evolution
	$\hat{b}_\pm(t)=e^{\pm\lambda t}\hat{b}_\pm(0)$.
	They are related to the prequench ladder operators of Eq.~\eqref{eq:h0} by
	\begin{equation}
	\hat{b}_\pm=\frac{\eh{\mp i\phi}\hat{a}+\eh{\pm i\phi}\hat{a}^\dagger}{\sqrt{2\sin 2\phi}}, 
	\qquad \hat{a}|k\rangle=\sqrt{k}|k-1\rangle,
	\label{eq:quadratures_and_ladder_operators}
	\end{equation}
	with $\phi=\tan^{-1}(\omega/\lambda)$, such that they are $\mathcal{O}(\heff^0)$ when applied to the noninteracting states with $k =\mathcal{O}(N^0)$, i.e., the states with quantum fluctuations $\langle\hat{b}_\pm^2\rangle=\mathcal{O}(\heff^0)$.
	
	With these definitions, one can formally use $\heff$ as a small parameter and the condition for the validity of the expansion \eqref{eq:Heisenberg_operator_expansion} is given by a \textit{local} Ehrenfest time
	\begin{equation}
	\heff\langle \bplus^2(t)\rangle \ll 1
	\quad 
	\Leftrightarrow \quad
	t\ll
	\frac{\log(\heff^{-1})}{2\lambda}
	\equiv t_E
	,
	\end{equation}
	characterized by the breakdown of the quadratic approximation.
	The expectation value can be taken in the ground state or in a thermal ensemble of the prequench system with temperature $k_BT/\Delta=\mathcal{O}(\heff^{0})$, where $\Delta $ is the single- or quasiparticle excitation energy.
	
	The explicit time dependence of $\hat{v}(t)$ takes the form
	\begin{equation}
	\hat{v}(t)=\sum_{\mu,\nu}v_{\mu\nu}\heff^{\frac{\mu+\nu}{2}}\eh{(\nu-\mu)\lambda t}\left\{\hat{b}_-^\mu \hat{b}_+^\nu\right\}_s,
	\label{eq:v_expansion}
	\end{equation}
	where the dependence on $\heff$ has been made also explicit and $\{\dots\}_s$ denotes symmetric ordering.
	The same can be done for $\hat{A}(t)$, assuming that the MF limit of $\hat{A}(0)$ is independent of $\heff$ \footnote{If $\hat{A}$ is a power series in $\heff$ the result applies to the $\heff$-independent coefficients individually.}, such that the time dependence (and the coefficients $v_{\mu\nu}$) can be pulled out of the commutators in Eq.~\eqref{eq:Heisenberg_operator_expansion} that is now organized as a power series in $\sqrt{\heff}$.
	By suppressing corrections of the form $t \eh{-\lambda t}$ for $t\gg 1/\lambda$ in the respective (operator valued) coefficients, one finds the dominant scaling \cite{SM}
	\begin{equation}
	\hat{A}_H(t)\sim \sum_{k}C_k\left(\sqrt{\heff}\eh{\lambda t}\bplus\right)^k,
	\label{eq:Heisenberg_operator_scaling}
	\end{equation}
	with the $\heff$-independent coefficients $C_k$ determined by the operators $\hat{A}$ and $\hat{v}$.
	Equation~\eqref{eq:quadratures_and_ladder_operators} implies
	\begin{equation}
	\langle k|\bplus^n|l\rangle =0 \qquad \text{for $n<|k-l|$},
	\end{equation}
	such that matrix elements scale as stated in Eq.~\eqref{eq:matrix_elements} with $c_{kl}=C_{|k-l|}\langle k|\bplus^{|k-l|} |l\rangle$ for $t\ll t_E$, with possible exceptions only if $C_{|k-l|}=0$ for specially constructed operators $\hat{A}$.
	
	Nonlinearities of the prequench system, Eq.~\eqref{eq:h} for $\alpha<1$, can now be included in a consistent way:
	The full eigenstates are expanded in a perturbation series around the harmonic oscillator states $|k\rangle$.
	One can then show using adiabatic switching \cite{SM} that these corrections do not contribute to the dominant scaling in Eq.~\eqref{eq:matrix_elements}, justifying the harmonic approximation \eqref{eq:h0}.
	Figure \ref{fig:two-site_BH_comparison} shows a comparison of the numerical simulations and the prediction in Eq.~\eqref{eq:matrix_elements} for $N=10^5$ particles and $\alpha=2.5$ using the operator $\hat{A}=\hat{z}+\hat{z}^2$.
	The fully analytical prediction (dashed lines) is clearly verified, as can be seen in the main plot.
	The top inset shows a collapse of the absolute values to the function $\heff\eh{2\lambda t}$ by calculating $\big|c_{kl}^{-1}\langle k|\hat{A}(t)|l\rangle\big|^{\frac{2}{|k-l|}}$.
	The bottom inset shows the deviations $\Delta \phi_{kl}=\arg\big[\eh{i(l-k)\phi}\langle k|\hat{A}(t)|l\rangle\big]$ between numerical and predicted phases (modulo $\pi$) accumulated in the time evolution, with $\phi$ defined in Eq.~\eqref{eq:quadratures_and_ladder_operators}.
	
	\textit{Expectation values and out-of-time-ordered correlators. --- }
	Although characterizing the time evolution of the off-diagonal matrix elements in a given basis solves the time-dependent problem, they cannot be observed directly. Nevertheless, a direct consequence of the universal form in  Eq.~\eqref{eq:matrix_elements} is that expectation values are given as power series in the parameters $\heff\eh{2\lambda t}$.
	If the system is in thermal equilibrium before the quench, this can be even refined using Wick's theorem in Eq.~\eqref{eq:Heisenberg_operator_scaling}, yielding
	\begin{equation}
	\big\langle\hat{A}(t)\big\rangle
	=
	\sum_m (2m-1)!C_{2m}\big(\heff\eh{2\lambda t}\big\langle\bplus^2\big\rangle\big)^m,
	\end{equation}
	with the temperature $k_BT=\beta^{-1}$ entering only via
	\begin{equation}
	\big\langle\bplus^2\big\rangle
	=
	\frac{\coth(\beta\Delta/2)}{2\sin2\phi},
	\end{equation}
	for $\beta\Delta=\mathcal{O}(\heff^0)$.
	Here, $\Delta$ is the single-particle level spacing in the harmonic approximation, thus suggesting the further renormalization   $\heff^{\mathrm{(r)}}(t,\beta)=\heff \eh{2\lambda t}\coth{(\beta\Delta/2)}$.
	
\begin{figure}
		\centering
		\includegraphics[width=\linewidth]{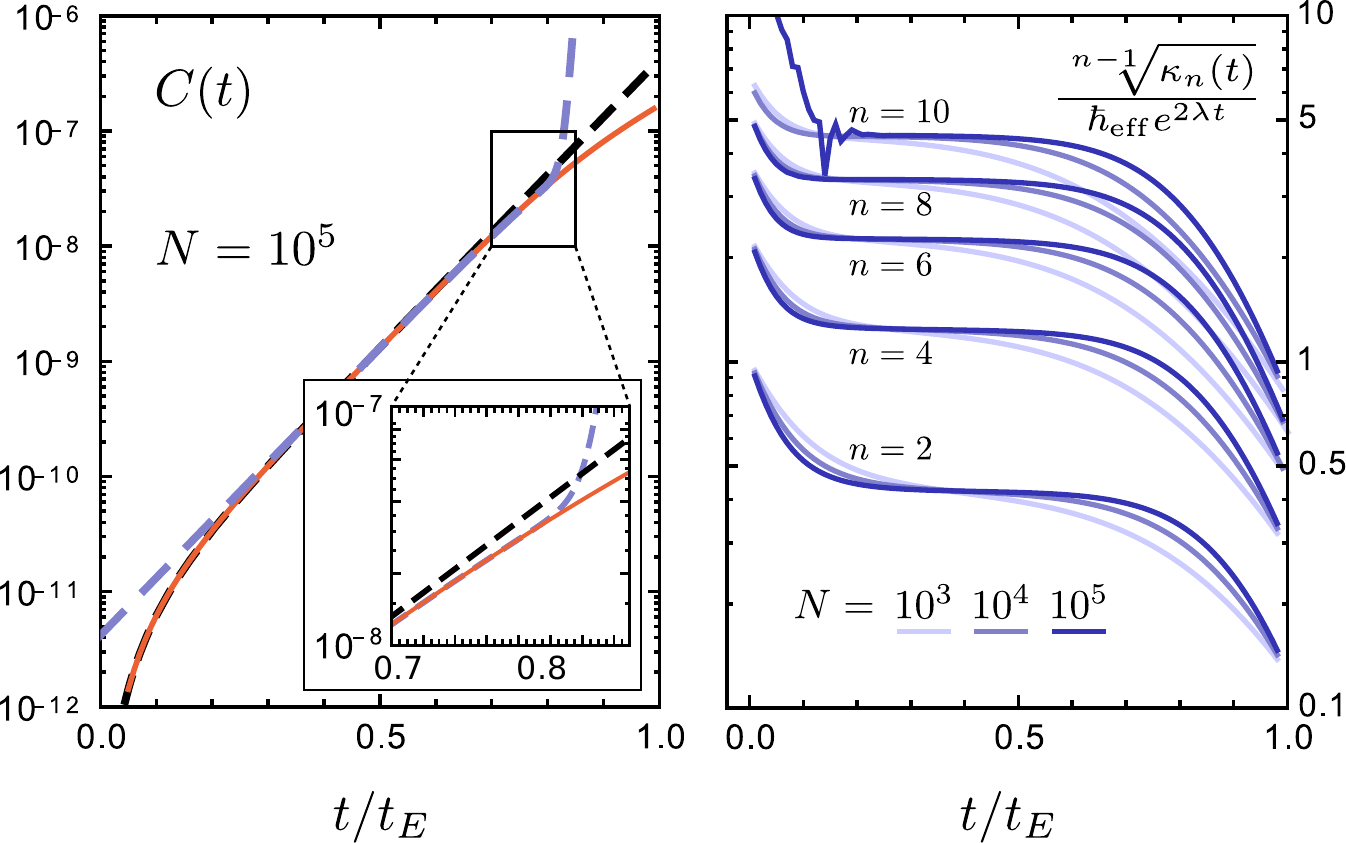}
		\caption{%
			Left: OTOC $C(t)=-\langle [\hat{A}(t),\hat{B}]^2\rangle$ (solid red) for $\hat{A}=\hat{B}=\hat{z}$ and the analytical prediction using the series expansion in the renormalized effective Planck constant (dashed blue).
			Dashed black lines show the analytic prediction for the leading orders including short-time corrections.
			Right: numerical check of the predicted behavior of the cumulants $\kappa_n(t)$, Eq.~\eqref{eq:cumulant_scaling}, for $n=2,4,6,8,10$ using roots.
			The oscillations for $n=10$ at short times are an artifact of the finite numerical precision.
			All data are for $\alpha=2.5$ with a prequench ($\alpha=0$) temperature $k_BT=2\Delta$.
		}
		\label{fig:OTOC_and_cumulants}
	\end{figure}

	A common probe for the instability properties in quantum systems is the out-of-time-ordered correlator (OTOC) \cite{Larkin1969,Rozenbaum2017,Cotler2018,Pappalardi2018,Lerose2020}.
	Using our approach, one can straightforwardly obtain a multiexponential form of the OTOC similar to a conjecture by Gu and Kitaev \cite{Gu2019},
	\begin{align}
	C(t)
	&=
	-\big\langle \big[\hat{A}(t),\hat{B}(0)\big]^2 \big\rangle 
	\nonumber
	\\&\sim
	\big(\heff \eh{\lambda t}\big)^2\sum_{m}c_{m}\big(\heff\eh{2\lambda t}\big\langle\bplus^{2}\big\rangle\big)^m,
	\label{eq:OTOC}
	\end{align}
	with ``classical'' coefficients $c_m$ determined by the constants $C_k$ in Eq.~\eqref{eq:Heisenberg_operator_scaling} and the linear expansion of $\hat{B}$ in $\bminus$, leading to the additional factor $\heff$ in Eq.~\eqref{eq:OTOC}.
	The OTOC scaling obtained from quasiclassical arguments \cite{Larkin1969},
	\begin{equation}
	C(t)\sim c_0\left(\heff\eh{\lambda t}\right)^2,
	\end{equation}
	corresponds only to the leading-order term $m=0$.
	One may now be tempted to use a finite number of terms in Eq.~\eqref{eq:OTOC} to obtain a better approximation at intermediate times.
	However, we report here the negative result that---at least for the integrable two-site Bose-Hubbard model---the corrections are very small within the region of convergence of the series, as can be seen in the left part of Fig.~\ref{fig:OTOC_and_cumulants}.
	There, the predicted series expansion in Eq.~\eqref{eq:OTOC} is plotted for a cutoff $m\leq 25$ (where $c_m=0$ for even $m$), showing a sharp breakdown at around $t\approx 0.8 t_E$, where the leading-order approximation (dashed black line) is still remarkably accurate.
	One should therefore not use higher-order exponentials for fitting data that do not show a clear exponential regime as was already noticed for the Sachdev-Ye-Kitaev model \cite{Kobrin2020}, a manifestation of the nonperturbative breakdown of the exponential behavior around $t_E$.
	
	One may, however, improve the prediction in the short-time regime $\lambda t\sim 1$ by using the full time dependence of the leading order in $\heff$, as is demonstrated by an analytic prediction \cite{SM} (dashed black line) in Fig.~\ref{fig:OTOC_and_cumulants}, accurate also at early times. This may allow for well-controlled fitting in many cases where $\heff=N^{-1}$ cannot be chosen arbitrarily small (cf.~\cite{Yao2016,Bohrdt2017,Shen2017,Pappalardi2018,Keles2019,Lantagne-Hurtubise2020}).
	
	The higher orders are, however, essential when it comes to cumulants of operators, as the $n$th cumulant of an operator of the form \eqref{eq:Heisenberg_operator_scaling} dominantly scales as \cite{SM} 
	\begin{equation}
	\kappa_n(t) \sim d_n \big(\heff\eh{2\lambda t}\big\langle\bplus^2\big\rangle\big)^{n-1},
	\label{eq:cumulant_scaling}
	\end{equation}
	with a constant $d_n$.
	Note that the $n$th \textit{moments} are expected to grow only as $(\heff\eh{2\lambda t})^{m}$ with $m=n/2$ or $m=(n+1)/2$ for even or odd $n$, such that various leading-order terms have to cancel in the cumulants for $n\geq 4$. Equation~\eqref{eq:cumulant_scaling} is verified in the right part of Fig.~\ref{fig:OTOC_and_cumulants}, where numerical results for the first five nonvanishing cumulants of $\hat{z}(t)$ are shown to follow such scaling while verifying $\heff \eh{2\lambda t}$ as the single relevant expansion parameter for this system.
	
	\textit{Generalization to more degrees of freedom. --- } Our analysis of the two-site model can be directly generalized to systems with more degrees of freedom that have a (symmetry protected) fixed point that undergoes a bifurcation at some critical coupling, yielding
	\begin{equation}
	\hat{A}(t)\sim \sum_{\boldsymbol{k}}C_{\boldsymbol{k}}\prod_i\left(\sqrt{\heff}\eh{\lambda_i t}\bplus^{(i)}\right)^{k_i},
	\end{equation}
	with index $i$ running through all the unstable directions characterized through $\bplus^{(i)}$ and the $\lambda_i$ are the respective MF divergence rates.
	The connection to the matrix elements is not as straightforward, in general, as the relation between the operators $\hat{b}_\pm^{(i)}$ and the bosonic operators characterizing the prequench stable dynamics can be any linear transformation, but the largest divergence rate will generically dominate the cumulants and expectation values.
	The Bose-Hubbard model with $L$ sites is special in this respect, as the interaction only enters as a quadratic term in the MF Hamiltonian, enabling a complete separation of the linearized dynamics that does not depend on the interaction \cite{SM}, allowing us also to write
	\begin{equation}
	\langle\boldsymbol{k}|\hat{A}(t)|\boldsymbol{l}\rangle \sim
	c_{\boldsymbol{kl}}\prod_{i}\big(\sqrt{\heff}\eh{\lambda_i t}\big)^{|k_i-l_i|}
	\label{eq:BH_generalized_matrix_elements}
	\end{equation}
	in the basis of the prequench eigenstates that is selected through an infinitesimal interaction.
	
	To show that the approach remains equally valid in systems without an integrable MF limit, a numerical simulation of the three-site Bose-Hubbard model
	\begin{equation}
	\hamiltonian=-J\sum_{j=1}^{3}(\cre{j}\ann{j+1}+\cre{j+1}\ann{j})+\frac{U}{2} \sum_{j=1}^{3}(\cre{j})^2(\ann{j})^2
	\end{equation}
	with periodic boundary conditions has been performed for $N=300$ particles and for an interaction quench from $U=0$ to $U=-20 J/N$.
	The MF analysis shows a bifurcation of the global energy minimum at $U=-9J/2N$ with two unstable directions having the same divergence rate $\lambda$ \cite{SM}.
	
	\begin{figure}
		\centering
		\includegraphics[width=\linewidth]{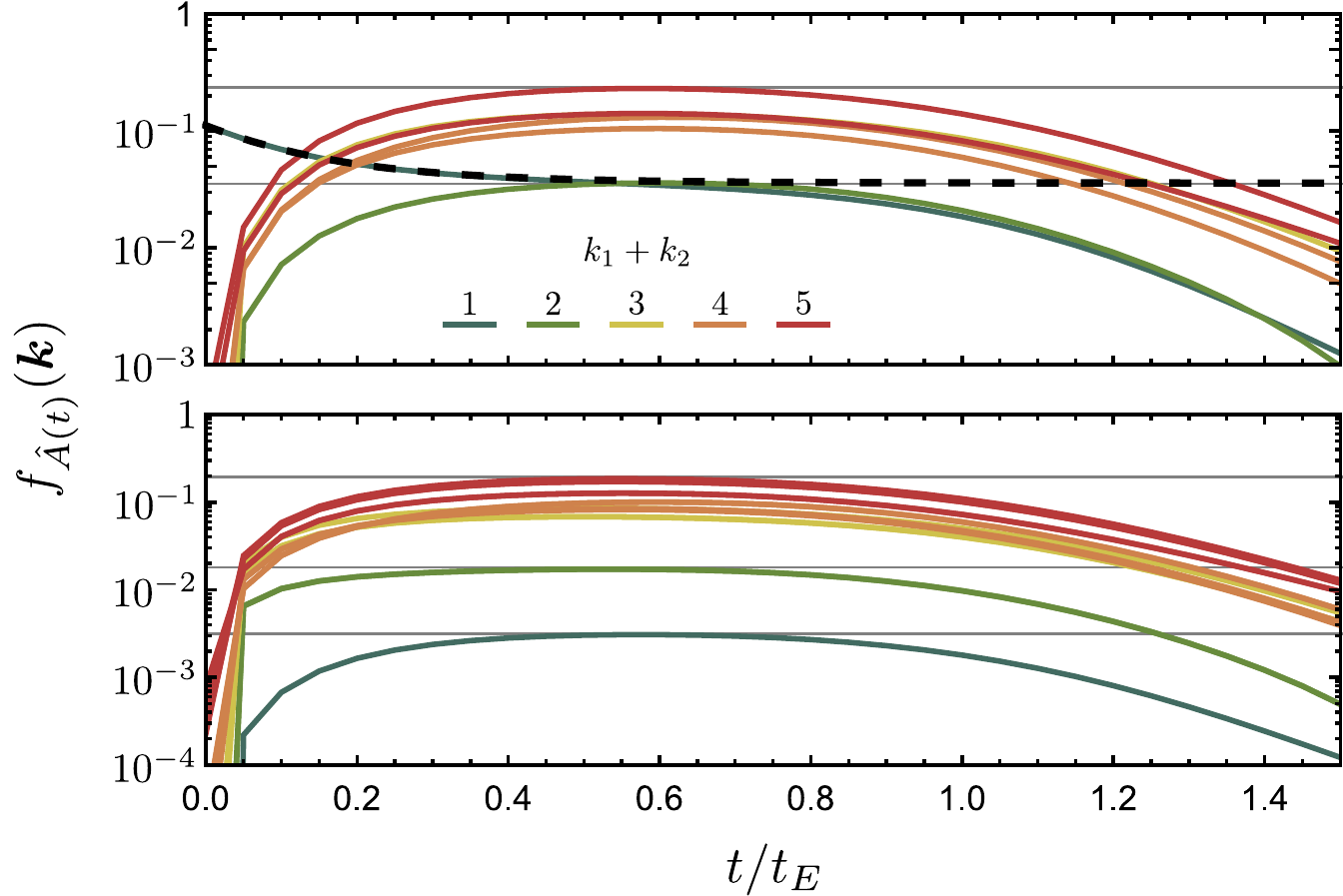}
		\caption{%
			Collapse of the absolute values of the nonvanishing matrix elements $\langle k_1k_2|\hat{A}(t)|0\rangle$ for $\hat{A}(t)=\hat{n}_1(t)/N$ (top) and $\hat{A}(t)=[\hat{n}_1(t),\hat{n}_2(0)]/N^{3/2}$ (bottom).
			Each color represents a value of the sum $k_1+k_2$ of the quantum numbers. 
			Thin gray horizontal lines representing the exponential predictions have been added to guide the eye.
			The dashed black line is the fully analytical prediction for $(k_1,k_2)=(1,0)$, showing that the short-time corrections to the asymptotic exponential form are still important for $N=300$ particles, but the tendency toward the predicted multiexponential behavior is clear and gets even more evident for the commutator (bottom).}
		\label{fig:3-site-BH}
	\end{figure}
	To verify the scaling of the matrix elements, Fig.~\ref{fig:3-site-BH} shows a collapse of the (nonvanishing) matrix elements to horizontal lines using the ansatz
	\begin{equation}
	f_{\hat{A}(t)}(\boldsymbol{k})
	=\frac{\big|\langle k_1k_2|\hat{A}(t)|0\rangle\big|^{\frac{2}{k_1+k_2}}}{\heff\eh{2\lambda t}}\sim \mathrm{const}
	\label{eq:matrix_elements_scaling_ansatz}
	\end{equation}
	for operators satisfying the scaling \eqref{eq:BH_generalized_matrix_elements}.
	The upper panel uses $\hat{A}(t)=\hat{z}_1(t)$,
	where $\hat{z}_i=\hat{n}_i/N$ is the (scaled) number operator on site $i$.
	The analytic prediction for $k_1+k_2=1$ that includes short-time corrections \cite{SM} is shown as a dashed black line and demonstrates that the deviations from exponential growth can be well explained by the fact that the requirement $t_E=\log N/2\lambda\gg\lambda^{-1}$ is not satisfied for $N=300$.
	The lower panel demonstrates that also the commutator of the number operators at different sites and different times has a scaling similar to Eq.~\eqref{eq:BH_generalized_matrix_elements}.
	For this, the ansatz \eqref{eq:matrix_elements_scaling_ansatz} with $\hat{A}(t)=[\hat{z}_1(t),\hat{z}_2(0)]/\sqrt{\heff}$
	is plotted for all excitations with $k_1+k_2\leq 5$.
	The factor $\heff^{-1/2}$ has been introduced to correct for $[\bplus^{(i)},\hat{z}_i]\propto \heff^{1/2}$.
	As can be seen in Fig.~\ref{fig:3-site-BH}, the individual curves clearly approach horizontal lines in a regime between $\lambda^{-1}\approx 0.35 t_E$ and $t_E$.
	
	\textit{Conclusions. --- }
	We have developed a general framework that shows how the relevant parameter for the dynamics after a quench across a critical point is given by a renormalized effective Planck constant $\eh{2\lambda t}/N$ in the quasiclassical regime $N\gg 1$ and for $\lambda t\gg1$.
	We support our analytical results by extensive numerical simulations for two exemplary critical scenarios: the self-trapping transition characteristic of integrable Josephson-like Hamiltonians and the nonintegrable three-site Bose-Hubbard model.
	While our approach is applicable to any system with bosonic order parameter in a well-defined mean-field limit, for these specific examples we uncover the predicted scaling in the matrix elements of generic operators.
	Although most observables are dominated by the first-order quasiclassical result, higher-order terms in the renormalized parameter are crucial when considering cumulants of simple operators, demonstrating the quantum nature of the results.
	
	Some results of this Letter on expectation values may also be obtained within approaches more suitable to address the approach to equilibrium at long times (e.g., Keldysh field theory).
	The method used here, however, directly addresses the short-time transient regime after a quench and is generally valid in any situation where a quantum-classical correspondence can be constructed.
	This may pave the way to generalize our results to the generic situation of chaotic or mixed (many-body) systems by restricting the dynamics to the vicinity of the less unstable classical periodic orbits \cite{Heller1984,Lerose2020} instead of stationary mean-field configurations as it was done here. This extension is the subject of present efforts.
	
	\begin{acknowledgments}
		We acknowledge funding through the Studien\-stiftung des Deutschen Volkes (BG) and the Deutsche Forschungsgemeinschaft through Project No.\ Ri681/14-1.
		We thank Q.~Hummel for useful conversations.
	\end{acknowledgments}
	
	\bibliography{../bibliography}

\begin{thebibliography}{61}%
\makeatletter
\providecommand \@ifxundefined [1]{%
 \@ifx{#1\undefined}
}%
\providecommand \@ifnum [1]{%
 \ifnum #1\expandafter \@firstoftwo
 \else \expandafter \@secondoftwo
 \fi
}%
\providecommand \@ifx [1]{%
 \ifx #1\expandafter \@firstoftwo
 \else \expandafter \@secondoftwo
 \fi
}%
\providecommand \natexlab [1]{#1}%
\providecommand \enquote  [1]{``#1''}%
\providecommand \bibnamefont  [1]{#1}%
\providecommand \bibfnamefont [1]{#1}%
\providecommand \citenamefont [1]{#1}%
\providecommand \href@noop [0]{\@secondoftwo}%
\providecommand \href [0]{\begingroup \@sanitize@url \@href}%
\providecommand \@href[1]{\@@startlink{#1}\@@href}%
\providecommand \@@href[1]{\endgroup#1\@@endlink}%
\providecommand \@sanitize@url [0]{\catcode `\\12\catcode `\$12\catcode
  `\&12\catcode `\#12\catcode `\^12\catcode `\_12\catcode `\%12\relax}%
\providecommand \@@startlink[1]{}%
\providecommand \@@endlink[0]{}%
\providecommand \url  [0]{\begingroup\@sanitize@url \@url }%
\providecommand \@url [1]{\endgroup\@href {#1}{\urlprefix }}%
\providecommand \urlprefix  [0]{URL }%
\providecommand \Eprint [0]{\href }%
\providecommand \doibase [0]{https://doi.org/}%
\providecommand \selectlanguage [0]{\@gobble}%
\providecommand \bibinfo  [0]{\@secondoftwo}%
\providecommand \bibfield  [0]{\@secondoftwo}%
\providecommand \translation [1]{[#1]}%
\providecommand \BibitemOpen [0]{}%
\providecommand \bibitemStop [0]{}%
\providecommand \bibitemNoStop [0]{.\EOS\space}%
\providecommand \EOS [0]{\spacefactor3000\relax}%
\providecommand \BibitemShut  [1]{\csname bibitem#1\endcsname}%
\let\auto@bib@innerbib\@empty
\bibitem [{\citenamefont {Wigner}(1932)}]{Wigner1932}%
  \BibitemOpen
  \bibfield  {author} {\bibinfo {author} {\bibfnamefont {E.}~\bibnamefont
  {Wigner}},\ }\bibfield  {title} {\bibinfo {title} {{On the quantum correction
  for thermodynamic equilibrium}},\ }\href
  {https://doi.org/10.1103/PhysRev.40.749} {\bibfield  {journal} {\bibinfo
  {journal} {Phys. Rev.}\ }\textbf {\bibinfo {volume} {40}},\ \bibinfo {pages}
  {749} (\bibinfo {year} {1932})}\BibitemShut {NoStop}%
\bibitem [{\citenamefont {Glauber}(2007)}]{Glauber2007}%
  \BibitemOpen
  \bibfield  {author} {\bibinfo {author} {\bibfnamefont {R.~J.}\ \bibnamefont
  {Glauber}},\ }\href@noop {} {\emph {\bibinfo {title} {{Quantum Theory of
  Optical Coherence}}}}\ (\bibinfo  {publisher} {Wiley-VCH Verlag},\ \bibinfo
  {address} {Berlin},\ \bibinfo {year} {2007})\BibitemShut {NoStop}%
\bibitem [{\citenamefont {Zurek}(2003)}]{Zurek2003}%
  \BibitemOpen
  \bibfield  {author} {\bibinfo {author} {\bibfnamefont {W.~H.}\ \bibnamefont
  {Zurek}},\ }\bibfield  {title} {\bibinfo {title} {{Decoherence, einselection,
  and the quantum origins of the classical}},\ }\href
  {https://doi.org/10.1103/RevModPhys.75.715} {\bibfield  {journal} {\bibinfo
  {journal} {Rev. Mod. Phys.}\ }\textbf {\bibinfo {volume} {75}},\ \bibinfo
  {pages} {715} (\bibinfo {year} {2003})}\BibitemShut {NoStop}%
\bibitem [{\citenamefont {Richter}(2000)}]{Richter2000}%
  \BibitemOpen
  \bibfield  {author} {\bibinfo {author} {\bibfnamefont {K.}~\bibnamefont
  {Richter}},\ }\href
  {http://www.physik.uni-regensburg.de/forschung/richter/richter/pages/research/springer-tracts-161.pdf}
  {\emph {\bibinfo {title} {{Semiclassical Theory of Mesoscopic Quantum
  Systems}}}},\ \bibinfo {series} {Springer Tracts in Modern Physics}, Vol.\
  \bibinfo {volume} {161}\ (\bibinfo  {publisher} {Springer-Verlag Berlin,
  Heidelberg},\ \bibinfo {year} {2000})\BibitemShut {NoStop}%
\bibitem [{\citenamefont {Brack}\ and\ \citenamefont
  {Bhaduri}(2003)}]{Brack2003}%
  \BibitemOpen
  \bibfield  {author} {\bibinfo {author} {\bibfnamefont {M.}~\bibnamefont
  {Brack}}\ and\ \bibinfo {author} {\bibfnamefont {R.~K.}\ \bibnamefont
  {Bhaduri}},\ }\href@noop {} {\emph {\bibinfo {title} {{Semiclassical
  Physics}}}},\ {Frontiers in Physics}\ (\bibinfo  {publisher} {Westview
  Press},\ \bibinfo {address} {Boulder},\ \bibinfo {year} {2003})\BibitemShut
  {NoStop}%
\bibitem [{\citenamefont {Hornberger}\ \emph {et~al.}(2012)\citenamefont
  {Hornberger}, \citenamefont {Gerlich}, \citenamefont {Haslinger},
  \citenamefont {Nimmrichter},\ and\ \citenamefont {Arndt}}]{Hornberger2012}%
  \BibitemOpen
  \bibfield  {author} {\bibinfo {author} {\bibfnamefont {K.}~\bibnamefont
  {Hornberger}}, \bibinfo {author} {\bibfnamefont {S.}~\bibnamefont {Gerlich}},
  \bibinfo {author} {\bibfnamefont {P.}~\bibnamefont {Haslinger}}, \bibinfo
  {author} {\bibfnamefont {S.}~\bibnamefont {Nimmrichter}},\ and\ \bibinfo
  {author} {\bibfnamefont {M.}~\bibnamefont {Arndt}},\ }\bibfield  {title}
  {\bibinfo {title} {{Colloquium: Quantum interference of clusters and
  molecules}},\ }\href {https://doi.org/10.1103/RevModPhys.84.157} {\bibfield
  {journal} {\bibinfo  {journal} {Rev. Mod. Phys.}\ }\textbf {\bibinfo {volume}
  {84}},\ \bibinfo {pages} {157} (\bibinfo {year} {2012})}\BibitemShut
  {NoStop}%
\bibitem [{\citenamefont {Polkovnikov}(2010)}]{Polkovnikov2010}%
  \BibitemOpen
  \bibfield  {author} {\bibinfo {author} {\bibfnamefont {A.}~\bibnamefont
  {Polkovnikov}},\ }\bibfield  {title} {\bibinfo {title} {{Phase space
  representation of quantum dynamics}},\ }\href
  {https://doi.org/10.1016/j.aop.2010.02.006} {\bibfield  {journal} {\bibinfo
  {journal} {Ann. Phys. (Amsterdam)}\ }\textbf {\bibinfo {volume} {325}},\
  \bibinfo {pages} {1790} (\bibinfo {year} {2010})}\BibitemShut {NoStop}%
\bibitem [{\citenamefont {Rundle}\ \emph {et~al.}(2019)\citenamefont {Rundle},
  \citenamefont {Tilma}, \citenamefont {Samson}, \citenamefont {Dwyer},
  \citenamefont {Bishop},\ and\ \citenamefont {Everitt}}]{Rundle2019}%
  \BibitemOpen
  \bibfield  {author} {\bibinfo {author} {\bibfnamefont {R.~P.}\ \bibnamefont
  {Rundle}}, \bibinfo {author} {\bibfnamefont {T.}~\bibnamefont {Tilma}},
  \bibinfo {author} {\bibfnamefont {J.~H.}\ \bibnamefont {Samson}}, \bibinfo
  {author} {\bibfnamefont {V.~M.}\ \bibnamefont {Dwyer}}, \bibinfo {author}
  {\bibfnamefont {R.~F.}\ \bibnamefont {Bishop}},\ and\ \bibinfo {author}
  {\bibfnamefont {M.~J.}\ \bibnamefont {Everitt}},\ }\bibfield  {title}
  {\bibinfo {title} {{General approach to quantum mechanics as a statistical
  theory}},\ }\href {https://doi.org/10.1103/PhysRevA.99.012115} {\bibfield
  {journal} {\bibinfo  {journal} {Phys. Rev. A}\ }\textbf {\bibinfo {volume}
  {99}},\ \bibinfo {pages} {012115} (\bibinfo {year} {2019})}\BibitemShut
  {NoStop}%
\bibitem [{\citenamefont {Sinatra}\ \emph {et~al.}(2002)\citenamefont
  {Sinatra}, \citenamefont {Lobo},\ and\ \citenamefont {Castin}}]{Sinatra2002}%
  \BibitemOpen
  \bibfield  {author} {\bibinfo {author} {\bibfnamefont {A.}~\bibnamefont
  {Sinatra}}, \bibinfo {author} {\bibfnamefont {C.}~\bibnamefont {Lobo}},\ and\
  \bibinfo {author} {\bibfnamefont {Y.}~\bibnamefont {Castin}},\ }\bibfield
  {title} {\bibinfo {title} {{The truncated Wigner method for Bose-condensed
  gases: Limits of validity and applications}},\ }\href
  {https://doi.org/10.1088/0953-4075/35/17/301} {\bibfield  {journal} {\bibinfo
   {journal} {J. Phys. B}\ }\textbf {\bibinfo {volume} {35}},\ \bibinfo {pages}
  {3599} (\bibinfo {year} {2002})}\BibitemShut {NoStop}%
\bibitem [{\citenamefont {Norrie}\ \emph {et~al.}(2006)\citenamefont {Norrie},
  \citenamefont {Ballagh},\ and\ \citenamefont {Gardiner}}]{Norrie2006}%
  \BibitemOpen
  \bibfield  {author} {\bibinfo {author} {\bibfnamefont {A.~A.}\ \bibnamefont
  {Norrie}}, \bibinfo {author} {\bibfnamefont {R.~J.}\ \bibnamefont
  {Ballagh}},\ and\ \bibinfo {author} {\bibfnamefont {C.~W.}\ \bibnamefont
  {Gardiner}},\ }\bibfield  {title} {\bibinfo {title} {{Quantum turbulence and
  correlations in Bose-Einstein condensate collisions}},\ }\href
  {https://doi.org/10.1103/PhysRevA.73.043617} {\bibfield  {journal} {\bibinfo
  {journal} {Phys. Rev. A}\ }\textbf {\bibinfo {volume} {73}},\ \bibinfo
  {pages} {043617} (\bibinfo {year} {2006})}\BibitemShut {NoStop}%
\bibitem [{\citenamefont {Deuar}\ and\ \citenamefont
  {Drummond}(2007)}]{Deuar2007}%
  \BibitemOpen
  \bibfield  {author} {\bibinfo {author} {\bibfnamefont {P.}~\bibnamefont
  {Deuar}}\ and\ \bibinfo {author} {\bibfnamefont {P.~D.}\ \bibnamefont
  {Drummond}},\ }\bibfield  {title} {\bibinfo {title} {{Correlations in a BEC
  Collision: First-Principles Quantum Dynamics with 150~000 Atoms}},\ }\href
  {https://doi.org/10.1103/PhysRevLett.98.120402} {\bibfield  {journal}
  {\bibinfo  {journal} {Phys. Rev. Lett.}\ }\textbf {\bibinfo {volume} {98}},\
  \bibinfo {pages} {120402} (\bibinfo {year} {2007})}\BibitemShut {NoStop}%
\bibitem [{\citenamefont {Blakie}\ \emph {et~al.}(2008)\citenamefont {Blakie},
  \citenamefont {Bradley}, \citenamefont {Davis}, \citenamefont {Ballagh},\
  and\ \citenamefont {Gardiner}}]{Blakie2008}%
  \BibitemOpen
  \bibfield  {author} {\bibinfo {author} {\bibfnamefont {P.}~\bibnamefont
  {Blakie}}, \bibinfo {author} {\bibfnamefont {A.}~\bibnamefont {Bradley}},
  \bibinfo {author} {\bibfnamefont {M.}~\bibnamefont {Davis}}, \bibinfo
  {author} {\bibfnamefont {R.}~\bibnamefont {Ballagh}},\ and\ \bibinfo {author}
  {\bibfnamefont {C.}~\bibnamefont {Gardiner}},\ }\bibfield  {title} {\bibinfo
  {title} {{Dynamics and statistical mechanics of ultra-cold Bose gases using
  $c$-field techniques}},\ }\href {https://doi.org/10.1080/00018730802564254}
  {\bibfield  {journal} {\bibinfo  {journal} {Adv. Phys.}\ }\textbf {\bibinfo
  {volume} {57}},\ \bibinfo {pages} {363} (\bibinfo {year} {2008})}\BibitemShut
  {NoStop}%
\bibitem [{\citenamefont {Fujiwara}\ \emph {et~al.}(1982)\citenamefont
  {Fujiwara}, \citenamefont {Osborn},\ and\ \citenamefont
  {Wilk}}]{Fujiwara1982}%
  \BibitemOpen
  \bibfield  {author} {\bibinfo {author} {\bibfnamefont {Y.}~\bibnamefont
  {Fujiwara}}, \bibinfo {author} {\bibfnamefont {T.~A.}\ \bibnamefont
  {Osborn}},\ and\ \bibinfo {author} {\bibfnamefont {S.~F.~J.}\ \bibnamefont
  {Wilk}},\ }\bibfield  {title} {\bibinfo {title} {{Wigner-Kirkwood
  expansions}},\ }\href {https://doi.org/10.1103/PhysRevA.25.14} {\bibfield
  {journal} {\bibinfo  {journal} {Phys. Rev. A}\ }\textbf {\bibinfo {volume}
  {25}},\ \bibinfo {pages} {14} (\bibinfo {year} {1982})}\BibitemShut {NoStop}%
\bibitem [{Note1()}]{Note1}%
  \BibitemOpen
  \bibinfo {note} {Quasiclassical refers to neglecting interfering classical
  solutions.}\BibitemShut {Stop}%
\bibitem [{\citenamefont {Gasenzer}\ \emph {et~al.}(2005)\citenamefont
  {Gasenzer}, \citenamefont {Berges}, \citenamefont {Schmidt},\ and\
  \citenamefont {Seco}}]{Gasenzer2005}%
  \BibitemOpen
  \bibfield  {author} {\bibinfo {author} {\bibfnamefont {T.}~\bibnamefont
  {Gasenzer}}, \bibinfo {author} {\bibfnamefont {J.}~\bibnamefont {Berges}},
  \bibinfo {author} {\bibfnamefont {M.}~\bibnamefont {Schmidt}},\ and\ \bibinfo
  {author} {\bibfnamefont {M.}~\bibnamefont {Seco}},\ }\bibfield  {title}
  {\bibinfo {title} {Nonperturbative dynamical many-body theory of a
  {Bose-Einstein} condensate},\ }\href
  {https://doi.org/10.1103/PhysRevA.72.063604} {\bibfield  {journal} {\bibinfo
  {journal} {Phys. Rev. A}\ }\textbf {\bibinfo {volume} {72}},\ \bibinfo
  {pages} {063604} (\bibinfo {year} {2005})}\BibitemShut {NoStop}%
\bibitem [{\citenamefont {Witthaut}\ \emph {et~al.}(2011)\citenamefont
  {Witthaut}, \citenamefont {Trimborn}, \citenamefont {Hennig}, \citenamefont
  {Kordas}, \citenamefont {Geisel},\ and\ \citenamefont
  {Wimberger}}]{Witthaut2011}%
  \BibitemOpen
  \bibfield  {author} {\bibinfo {author} {\bibfnamefont {D.}~\bibnamefont
  {Witthaut}}, \bibinfo {author} {\bibfnamefont {F.}~\bibnamefont {Trimborn}},
  \bibinfo {author} {\bibfnamefont {H.}~\bibnamefont {Hennig}}, \bibinfo
  {author} {\bibfnamefont {G.}~\bibnamefont {Kordas}}, \bibinfo {author}
  {\bibfnamefont {T.}~\bibnamefont {Geisel}},\ and\ \bibinfo {author}
  {\bibfnamefont {S.}~\bibnamefont {Wimberger}},\ }\bibfield  {title} {\bibinfo
  {title} {Beyond mean-field dynamics in open {Bose-Hubbard} chains},\ }\href
  {https://doi.org/10.1103/PhysRevA.83.063608} {\bibfield  {journal} {\bibinfo
  {journal} {Phys. Rev. A}\ }\textbf {\bibinfo {volume} {83}},\ \bibinfo
  {pages} {063608} (\bibinfo {year} {2011})}\BibitemShut {NoStop}%
\bibitem [{\citenamefont {Kordas}\ \emph {et~al.}(2013)\citenamefont {Kordas},
  \citenamefont {Wimberger},\ and\ \citenamefont {Witthaut}}]{Kordas2013}%
  \BibitemOpen
  \bibfield  {author} {\bibinfo {author} {\bibfnamefont {G.}~\bibnamefont
  {Kordas}}, \bibinfo {author} {\bibfnamefont {S.}~\bibnamefont {Wimberger}},\
  and\ \bibinfo {author} {\bibfnamefont {D.}~\bibnamefont {Witthaut}},\
  }\bibfield  {title} {\bibinfo {title} {Decay and fragmentation in an open
  {Bose-Hubbard} chain},\ }\href {https://doi.org/10.1103/PhysRevA.87.043618}
  {\bibfield  {journal} {\bibinfo  {journal} {Phys. Rev. A}\ }\textbf {\bibinfo
  {volume} {87}},\ \bibinfo {pages} {043618} (\bibinfo {year}
  {2013})}\BibitemShut {NoStop}%
\bibitem [{\citenamefont {Ivanov}\ \emph {et~al.}(2013)\citenamefont {Ivanov},
  \citenamefont {Kordas}, \citenamefont {Komnik},\ and\ \citenamefont
  {Wimberger}}]{Ivanov2013}%
  \BibitemOpen
  \bibfield  {author} {\bibinfo {author} {\bibfnamefont {A.}~\bibnamefont
  {Ivanov}}, \bibinfo {author} {\bibfnamefont {G.}~\bibnamefont {Kordas}},
  \bibinfo {author} {\bibfnamefont {A.}~\bibnamefont {Komnik}},\ and\ \bibinfo
  {author} {\bibfnamefont {S.}~\bibnamefont {Wimberger}},\ }\bibfield  {title}
  {\bibinfo {title} {Bosonic transport through a chain of quantum dots},\
  }\href {https://doi.org/10.1140/epjb/e2013-40417-4} {\bibfield  {journal}
  {\bibinfo  {journal} {Eur. Phys. J. B}\ }\textbf {\bibinfo {volume} {86}},\
  \bibinfo {pages} {345} (\bibinfo {year} {2013})}\BibitemShut {NoStop}%
\bibitem [{\citenamefont {Chatterjee}(1990)}]{Chatterjee1990}%
  \BibitemOpen
  \bibfield  {author} {\bibinfo {author} {\bibfnamefont {A.}~\bibnamefont
  {Chatterjee}},\ }\bibfield  {title} {\bibinfo {title} {{Large-$N$ expansions
  in quantum mechanics, atomic physics and some O($N$) invariant systems}},\
  }\href {https://doi.org/10.1016/0370-1573(90)90048-7} {\bibfield  {journal}
  {\bibinfo  {journal} {Phys. Rep.}\ }\textbf {\bibinfo {volume} {186}},\
  \bibinfo {pages} {249} (\bibinfo {year} {1990})}\BibitemShut {NoStop}%
\bibitem [{\citenamefont {Zibold}\ \emph {et~al.}(2010)\citenamefont {Zibold},
  \citenamefont {Nicklas}, \citenamefont {Gross},\ and\ \citenamefont
  {Oberthaler}}]{Zibold2010}%
  \BibitemOpen
  \bibfield  {author} {\bibinfo {author} {\bibfnamefont {T.}~\bibnamefont
  {Zibold}}, \bibinfo {author} {\bibfnamefont {E.}~\bibnamefont {Nicklas}},
  \bibinfo {author} {\bibfnamefont {C.}~\bibnamefont {Gross}},\ and\ \bibinfo
  {author} {\bibfnamefont {M.~K.}\ \bibnamefont {Oberthaler}},\ }\bibfield
  {title} {\bibinfo {title} {{Classical Bifurcation at the Transition from Rabi
  to Josephson Dynamics}},\ }\href
  {https://doi.org/10.1103/PhysRevLett.105.204101} {\bibfield  {journal}
  {\bibinfo  {journal} {Phys. Rev. Lett.}\ }\textbf {\bibinfo {volume} {105}},\
  \bibinfo {pages} {204101} (\bibinfo {year} {2010})}\BibitemShut {NoStop}%
\bibitem [{\citenamefont {Polkovnikov}\ \emph {et~al.}(2011)\citenamefont
  {Polkovnikov}, \citenamefont {Sengupta}, \citenamefont {Silva},\ and\
  \citenamefont {Vengalattore}}]{Polkovnikov2011}%
  \BibitemOpen
  \bibfield  {author} {\bibinfo {author} {\bibfnamefont {A.}~\bibnamefont
  {Polkovnikov}}, \bibinfo {author} {\bibfnamefont {K.}~\bibnamefont
  {Sengupta}}, \bibinfo {author} {\bibfnamefont {A.}~\bibnamefont {Silva}},\
  and\ \bibinfo {author} {\bibfnamefont {M.}~\bibnamefont {Vengalattore}},\
  }\bibfield  {title} {\bibinfo {title} {{Colloquium: Nonequilibrium dynamics
  of closed interacting quantum systems}},\ }\href
  {https://doi.org/10.1103/RevModPhys.83.863} {\bibfield  {journal} {\bibinfo
  {journal} {Rev. Mod. Phys.}\ }\textbf {\bibinfo {volume} {83}},\ \bibinfo
  {pages} {863} (\bibinfo {year} {2011})}\BibitemShut {NoStop}%
\bibitem [{\citenamefont {Pappalardi}\ \emph {et~al.}(2018)\citenamefont
  {Pappalardi}, \citenamefont {Russomanno}, \citenamefont {{\ifmmode
  \check{Z}\else \v{Z}\fi{}unkovi\ifmmode \check{c}\else \v{c}\fi{}}},
  \citenamefont {Iemini}, \citenamefont {Silva},\ and\ \citenamefont
  {Fazio}}]{Pappalardi2018}%
  \BibitemOpen
  \bibfield  {author} {\bibinfo {author} {\bibfnamefont {S.}~\bibnamefont
  {Pappalardi}}, \bibinfo {author} {\bibfnamefont {A.}~\bibnamefont
  {Russomanno}}, \bibinfo {author} {\bibfnamefont {B.}~\bibnamefont {{\ifmmode
  \check{Z}\else \v{Z}\fi{}unkovi\ifmmode \check{c}\else \v{c}\fi{}}}},
  \bibinfo {author} {\bibfnamefont {F.}~\bibnamefont {Iemini}}, \bibinfo
  {author} {\bibfnamefont {A.}~\bibnamefont {Silva}},\ and\ \bibinfo {author}
  {\bibfnamefont {R.}~\bibnamefont {Fazio}},\ }\bibfield  {title} {\bibinfo
  {title} {{Scrambling and entanglement spreading in long-range spin chains}},\
  }\href {https://doi.org/10.1103/PhysRevB.98.134303} {\bibfield  {journal}
  {\bibinfo  {journal} {Phys. Rev. B}\ }\textbf {\bibinfo {volume} {98}},\
  \bibinfo {pages} {134303} (\bibinfo {year} {2018})}\BibitemShut {NoStop}%
\bibitem [{\citenamefont {Mitra}(2018)}]{Mitra2018}%
  \BibitemOpen
  \bibfield  {author} {\bibinfo {author} {\bibfnamefont {A.}~\bibnamefont
  {Mitra}},\ }\bibfield  {title} {\bibinfo {title} {{Quantum quench
  dynamics}},\ }\href
  {https://doi.org/10.1146/annurev-conmatphys-031016-025451} {\bibfield
  {journal} {\bibinfo  {journal} {Annu. Rev. Condens. Matter Phys.}\ }\textbf
  {\bibinfo {volume} {9}},\ \bibinfo {pages} {245} (\bibinfo {year}
  {2018})}\BibitemShut {NoStop}%
\bibitem [{\citenamefont {Emary}\ and\ \citenamefont
  {Brandes}(2003)}]{Emary2003}%
  \BibitemOpen
  \bibfield  {author} {\bibinfo {author} {\bibfnamefont {C.}~\bibnamefont
  {Emary}}\ and\ \bibinfo {author} {\bibfnamefont {T.}~\bibnamefont
  {Brandes}},\ }\bibfield  {title} {\bibinfo {title} {{Chaos and the quantum
  phase transition in the Dicke model}},\ }\href
  {https://doi.org/10.1103/PhysRevE.67.066203} {\bibfield  {journal} {\bibinfo
  {journal} {Phys. Rev. E}\ }\textbf {\bibinfo {volume} {67}},\ \bibinfo
  {pages} {066203} (\bibinfo {year} {2003})}\BibitemShut {NoStop}%
\bibitem [{\citenamefont {Caprio}\ \emph {et~al.}(2008)\citenamefont {Caprio},
  \citenamefont {Cejnar},\ and\ \citenamefont {Iachello}}]{Caprio2008}%
  \BibitemOpen
  \bibfield  {author} {\bibinfo {author} {\bibfnamefont {M.~A.}\ \bibnamefont
  {Caprio}}, \bibinfo {author} {\bibfnamefont {P.}~\bibnamefont {Cejnar}},\
  and\ \bibinfo {author} {\bibfnamefont {F.}~\bibnamefont {Iachello}},\
  }\bibfield  {title} {\bibinfo {title} {{Excited state quantum phase
  transitions in many-body systems}},\ }\href
  {https://doi.org/10.1016/j.aop.2007.06.011} {\bibfield  {journal} {\bibinfo
  {journal} {Ann. Phys. (Amsterdam)}\ }\textbf {\bibinfo {volume} {323}},\
  \bibinfo {pages} {1106} (\bibinfo {year} {2008})}\BibitemShut {NoStop}%
\bibitem [{\citenamefont {Bastidas}\ \emph {et~al.}(2014)\citenamefont
  {Bastidas}, \citenamefont {P{\'e}rez-Fern{\'a}ndez}, \citenamefont {Vogl},\
  and\ \citenamefont {Brandes}}]{Bastidas2014}%
  \BibitemOpen
  \bibfield  {author} {\bibinfo {author} {\bibfnamefont {V.~M.}\ \bibnamefont
  {Bastidas}}, \bibinfo {author} {\bibfnamefont {P.}~\bibnamefont
  {P{\'e}rez-Fern{\'a}ndez}}, \bibinfo {author} {\bibfnamefont
  {M.}~\bibnamefont {Vogl}},\ and\ \bibinfo {author} {\bibfnamefont
  {T.}~\bibnamefont {Brandes}},\ }\bibfield  {title} {\bibinfo {title}
  {{Quantum Criticality and Dynamical Instability in the Kicked-Top Model}},\
  }\href {https://doi.org/10.1103/PhysRevLett.112.140408} {\bibfield  {journal}
  {\bibinfo  {journal} {Phys. Rev. Lett.}\ }\textbf {\bibinfo {volume} {112}},\
  \bibinfo {pages} {140408} (\bibinfo {year} {2014})}\BibitemShut {NoStop}%
\bibitem [{\citenamefont {Str{\'a}nsk{\'y}}\ \emph {et~al.}(2014)\citenamefont
  {Str{\'a}nsk{\'y}}, \citenamefont {Macek},\ and\ \citenamefont
  {Cejnar}}]{Stransky2014}%
  \BibitemOpen
  \bibfield  {author} {\bibinfo {author} {\bibfnamefont {P.}~\bibnamefont
  {Str{\'a}nsk{\'y}}}, \bibinfo {author} {\bibfnamefont {M.}~\bibnamefont
  {Macek}},\ and\ \bibinfo {author} {\bibfnamefont {P.}~\bibnamefont
  {Cejnar}},\ }\bibfield  {title} {\bibinfo {title} {{Excited-state quantum
  phase transitions in systems with two degrees of freedom: Level density,
  level dynamics, thermal properties}},\ }\href
  {https://doi.org/10.1016/j.aop.2014.03.006} {\bibfield  {journal} {\bibinfo
  {journal} {Ann. Phys. (Amsterdam)}\ }\textbf {\bibinfo {volume} {345}},\
  \bibinfo {pages} {73} (\bibinfo {year} {2014})}\BibitemShut {NoStop}%
\bibitem [{\citenamefont {Bastarrachea-Magnani}\ \emph
  {et~al.}(2016)\citenamefont {Bastarrachea-Magnani}, \citenamefont
  {Lerma-Hern{\'a}ndez},\ and\ \citenamefont
  {Hirsch}}]{Bastarrachea-Magnani2016}%
  \BibitemOpen
  \bibfield  {author} {\bibinfo {author} {\bibfnamefont {M.~A.}\ \bibnamefont
  {Bastarrachea-Magnani}}, \bibinfo {author} {\bibfnamefont {S.}~\bibnamefont
  {Lerma-Hern{\'a}ndez}},\ and\ \bibinfo {author} {\bibfnamefont {J.~G.}\
  \bibnamefont {Hirsch}},\ }\bibfield  {title} {\bibinfo {title} {{Thermal and
  quantum phase transitions in atom-field systems: A microcanonical
  analysis}},\ }\href {http://stacks.iop.org/1742-5468/2016/i=9/a=093105}
  {\bibfield  {journal} {\bibinfo  {journal} {J. Stat. Mech.}\ }\textbf
  {\bibinfo {volume} {2016}},\ \bibinfo {pages} {093105} (\bibinfo {year}
  {2016})}\BibitemShut {NoStop}%
\bibitem [{\citenamefont {Rubeni}\ \emph {et~al.}(2017)\citenamefont {Rubeni},
  \citenamefont {Links}, \citenamefont {Isaac},\ and\ \citenamefont
  {Foerster}}]{Rubeni2017}%
  \BibitemOpen
  \bibfield  {author} {\bibinfo {author} {\bibfnamefont {D.}~\bibnamefont
  {Rubeni}}, \bibinfo {author} {\bibfnamefont {J.}~\bibnamefont {Links}},
  \bibinfo {author} {\bibfnamefont {P.~S.}\ \bibnamefont {Isaac}},\ and\
  \bibinfo {author} {\bibfnamefont {A.}~\bibnamefont {Foerster}},\ }\bibfield
  {title} {\bibinfo {title} {{Two-site Bose-Hubbard model with nonlinear
  tunneling: Classical and quantum analysis}},\ }\href
  {https://doi.org/10.1103/PhysRevA.95.043607} {\bibfield  {journal} {\bibinfo
  {journal} {Phys. Rev. A}\ }\textbf {\bibinfo {volume} {95}},\ \bibinfo
  {pages} {043607} (\bibinfo {year} {2017})}\BibitemShut {NoStop}%
\bibitem [{\citenamefont {Hummel}\ \emph {et~al.}(2019)\citenamefont {Hummel},
  \citenamefont {Geiger}, \citenamefont {Urbina},\ and\ \citenamefont
  {Richter}}]{Hummel2019b}%
  \BibitemOpen
  \bibfield  {author} {\bibinfo {author} {\bibfnamefont {Q.}~\bibnamefont
  {Hummel}}, \bibinfo {author} {\bibfnamefont {B.}~\bibnamefont {Geiger}},
  \bibinfo {author} {\bibfnamefont {J.~D.}\ \bibnamefont {Urbina}},\ and\
  \bibinfo {author} {\bibfnamefont {K.}~\bibnamefont {Richter}},\ }\bibfield
  {title} {\bibinfo {title} {{Reversible Quantum Information Spreading in
  Many-Body Systems near Criticality}},\ }\href
  {https://doi.org/10.1103/PhysRevLett.123.160401} {\bibfield  {journal}
  {\bibinfo  {journal} {Phys. Rev. Lett.}\ }\textbf {\bibinfo {volume} {123}},\
  \bibinfo {pages} {160401} (\bibinfo {year} {2019})}\BibitemShut {NoStop}%
\bibitem [{\citenamefont {Gutzwiller}(1990)}]{Gutzwiller1990}%
  \BibitemOpen
  \bibfield  {author} {\bibinfo {author} {\bibfnamefont {M.~C.}\ \bibnamefont
  {Gutzwiller}},\ }\href@noop {} {\emph {\bibinfo {title} {{Chaos in Classical
  and Quantum Mechanics}}}},\ {Interdisciplinary Applied Mathematics}\
  (\bibinfo  {publisher} {Springer},\ \bibinfo {address} {New York},\ \bibinfo
  {year} {1990})\BibitemShut {NoStop}%
\bibitem [{\citenamefont {Chirikov}\ \emph {et~al.}(1981)\citenamefont
  {Chirikov}, \citenamefont {Izrailev},\ and\ \citenamefont
  {Shepelyansky}}]{Chirikov1981}%
  \BibitemOpen
  \bibfield  {author} {\bibinfo {author} {\bibfnamefont {B.~V.}\ \bibnamefont
  {Chirikov}}, \bibinfo {author} {\bibfnamefont {F.}~\bibnamefont {Izrailev}},\
  and\ \bibinfo {author} {\bibfnamefont {D.}~\bibnamefont {Shepelyansky}},\
  }\bibfield  {title} {\bibinfo {title} {Dynamical stochasticity in classical
  and quantum mechanics},\ }\href
  {https://www.quantware.ups-tlse.fr/chirikov/refs/chi1981a.pdf} {\bibfield
  {journal} {\bibinfo  {journal} {Sov. Sci. Rev. C}\ }\textbf {\bibinfo
  {volume} {2}},\ \bibinfo {pages} {209} (\bibinfo {year} {1981})}\BibitemShut
  {NoStop}%
\bibitem [{\citenamefont {Rammensee}\ \emph {et~al.}(2018)\citenamefont
  {Rammensee}, \citenamefont {Urbina},\ and\ \citenamefont
  {Richter}}]{Rammensee2018}%
  \BibitemOpen
  \bibfield  {author} {\bibinfo {author} {\bibfnamefont {J.}~\bibnamefont
  {Rammensee}}, \bibinfo {author} {\bibfnamefont {J.~D.}\ \bibnamefont
  {Urbina}},\ and\ \bibinfo {author} {\bibfnamefont {K.}~\bibnamefont
  {Richter}},\ }\bibfield  {title} {\bibinfo {title} {{Many-Body Quantum
  Interference and the Saturation of Out-of-Time-Order Correlators}},\ }\href
  {https://doi.org/10.1103/PhysRevLett.121.124101} {\bibfield  {journal}
  {\bibinfo  {journal} {Phys. Rev. Lett.}\ }\textbf {\bibinfo {volume} {121}},\
  \bibinfo {pages} {124101} (\bibinfo {year} {2018})}\BibitemShut {NoStop}%
\bibitem [{SM()}]{SM}%
  \BibitemOpen
  \href@noop {} {}\bibinfo {note} {See Supplemental Material at
  \url{https://arxiv.org/abs/2010.08364}, Sec.~A for the mean-field analysis of
  the Bose-Hubbard (BH) dimer, Sec.~B for the proof of
  Eq.~\eqref{eq:Heisenberg_operator_scaling} and the coefficients $C_k$ and
  $c_{kl}$, Sec.~C for the proof of subdominance of perturbations in the
  states, Sec.~D for the analysis of the BH model with $L$ sites, and Sec.~E
  for the explicit short-time first-order results used in Figs.~2 and 3.
  Includes additional
  Refs.~\cite{Jasiulewicz2003,James1962,Fetter1971}.}\BibitemShut {Stop}%
\bibitem [{\citenamefont {Jasiulewicz}\ and\ \citenamefont
  {Kordecki}(2003)}]{Jasiulewicz2003}%
  \BibitemOpen
  \bibfield  {author} {\bibinfo {author} {\bibfnamefont {H.}~\bibnamefont
  {Jasiulewicz}}\ and\ \bibinfo {author} {\bibfnamefont {W.}~\bibnamefont
  {Kordecki}},\ }\bibfield  {title} {\bibinfo {title} {{C}onvolutions of
  {E}rlang and of {P}ascal distributions with applications to reliability},\
  }\href {https://doi.org/10.1515/dema-2003-0125} {\bibfield  {journal}
  {\bibinfo  {journal} {Demonstr. Math.}\ }\textbf {\bibinfo {volume} {36}},\
  \bibinfo {pages} {231} (\bibinfo {year} {2003})}\BibitemShut {NoStop}%
\bibitem [{\citenamefont {James}\ and\ \citenamefont
  {Mayne}(1962)}]{James1962}%
  \BibitemOpen
  \bibfield  {author} {\bibinfo {author} {\bibfnamefont {G.~S.}\ \bibnamefont
  {James}}\ and\ \bibinfo {author} {\bibfnamefont {A.~J.}\ \bibnamefont
  {Mayne}},\ }\bibfield  {title} {\bibinfo {title} {{Cumulants of functions of
  random variables}},\ }\href {https://about.jstor.org/terms
  http://www.jstor.org/stable/25049192} {\bibfield  {journal} {\bibinfo
  {journal} {Sankhy Ser. A (1961-2002)}\ }\textbf {\bibinfo {volume} {24}},\
  \bibinfo {pages} {47} (\bibinfo {year} {1962})}\BibitemShut {NoStop}%
\bibitem [{\citenamefont {Fetter}\ and\ \citenamefont
  {Walecka}(1971)}]{Fetter1971}%
  \BibitemOpen
  \bibfield  {author} {\bibinfo {author} {\bibfnamefont {A.~L.}\ \bibnamefont
  {Fetter}}\ and\ \bibinfo {author} {\bibfnamefont {J.~D.}\ \bibnamefont
  {Walecka}},\ }\href@noop {} {\emph {\bibinfo {title} {{Quantum Theory of
  Many-Particle Systems}}}},\ International Series in Pure and Appled Physics\
  (\bibinfo  {publisher} {McGraw-Hill Book Company},\ \bibinfo {address} {New
  York},\ \bibinfo {year} {1971})\BibitemShut {NoStop}%
\bibitem [{\citenamefont {Juli{\'{a}}-D{\'{i}}az}\ \emph
  {et~al.}(2010)\citenamefont {Juli{\'{a}}-D{\'{i}}az}, \citenamefont
  {Dagnino}, \citenamefont {Lewenstein}, \citenamefont {Martorell},\ and\
  \citenamefont {Polls}}]{Julia-Diaz2010}%
  \BibitemOpen
  \bibfield  {author} {\bibinfo {author} {\bibfnamefont {B.}~\bibnamefont
  {Juli{\'{a}}-D{\'{i}}az}}, \bibinfo {author} {\bibfnamefont {D.}~\bibnamefont
  {Dagnino}}, \bibinfo {author} {\bibfnamefont {M.}~\bibnamefont {Lewenstein}},
  \bibinfo {author} {\bibfnamefont {J.}~\bibnamefont {Martorell}},\ and\
  \bibinfo {author} {\bibfnamefont {A.}~\bibnamefont {Polls}},\ }\bibfield
  {title} {\bibinfo {title} {{Macroscopic self-trapping in Bose-Einstein
  condensates: Analysis of a dynamical quantum phase transition}},\ }\href
  {https://doi.org/10.1103/PhysRevA.81.023615} {\bibfield  {journal} {\bibinfo
  {journal} {Phys. Rev. A}\ }\textbf {\bibinfo {volume} {81}},\ \bibinfo
  {pages} {023615} (\bibinfo {year} {2010})}\BibitemShut {NoStop}%
\bibitem [{\citenamefont {Rautenberg}\ and\ \citenamefont
  {G{\"a}rttner}(2020)}]{Rautenberg2019}%
  \BibitemOpen
  \bibfield  {author} {\bibinfo {author} {\bibfnamefont {M.}~\bibnamefont
  {Rautenberg}}\ and\ \bibinfo {author} {\bibfnamefont {M.}~\bibnamefont
  {G{\"a}rttner}},\ }\bibfield  {title} {\bibinfo {title} {{Classical and
  quantum chaos in a three-mode bosonic system}},\ }\href
  {https://doi.org/10.1103/PhysRevA.101.053604} {\bibfield  {journal} {\bibinfo
   {journal} {Phys. Rev. A}\ }\textbf {\bibinfo {volume} {101}},\ \bibinfo
  {pages} {053604} (\bibinfo {year} {2020})}\BibitemShut {NoStop}%
\bibitem [{\citenamefont {Kunkel}\ \emph {et~al.}(2019)\citenamefont {Kunkel},
  \citenamefont {Pr{\"{u}}fer}, \citenamefont {Lannig}, \citenamefont
  {Rosa-Medina}, \citenamefont {Bonnin}, \citenamefont {G{\"{a}}rttner},
  \citenamefont {Strobel},\ and\ \citenamefont {Oberthaler}}]{Kunkel2019}%
  \BibitemOpen
  \bibfield  {author} {\bibinfo {author} {\bibfnamefont {P.}~\bibnamefont
  {Kunkel}}, \bibinfo {author} {\bibfnamefont {M.}~\bibnamefont
  {Pr{\"{u}}fer}}, \bibinfo {author} {\bibfnamefont {S.}~\bibnamefont
  {Lannig}}, \bibinfo {author} {\bibfnamefont {R.}~\bibnamefont {Rosa-Medina}},
  \bibinfo {author} {\bibfnamefont {A.}~\bibnamefont {Bonnin}}, \bibinfo
  {author} {\bibfnamefont {M.}~\bibnamefont {G{\"{a}}rttner}}, \bibinfo
  {author} {\bibfnamefont {H.}~\bibnamefont {Strobel}},\ and\ \bibinfo {author}
  {\bibfnamefont {M.~K.}\ \bibnamefont {Oberthaler}},\ }\bibfield  {title}
  {\bibinfo {title} {{Simultaneous Readout of Noncommuting Collective Spin
  Observables beyond the Standard Quantum Limit}},\ }\href
  {https://doi.org/10.1103/PhysRevLett.123.063603} {\bibfield  {journal}
  {\bibinfo  {journal} {Phys. Rev. Lett.}\ }\textbf {\bibinfo {volume} {123}},\
  \bibinfo {pages} {063603} (\bibinfo {year} {2019})}\BibitemShut {NoStop}%
\bibitem [{\citenamefont {Nemoto}\ \emph {et~al.}(2000)\citenamefont {Nemoto},
  \citenamefont {Holmes}, \citenamefont {Milburn},\ and\ \citenamefont
  {Munro}}]{Nemoto2000}%
  \BibitemOpen
  \bibfield  {author} {\bibinfo {author} {\bibfnamefont {K.}~\bibnamefont
  {Nemoto}}, \bibinfo {author} {\bibfnamefont {C.~A.}\ \bibnamefont {Holmes}},
  \bibinfo {author} {\bibfnamefont {G.~J.}\ \bibnamefont {Milburn}},\ and\
  \bibinfo {author} {\bibfnamefont {W.~J.}\ \bibnamefont {Munro}},\ }\bibfield
  {title} {\bibinfo {title} {{Quantum dynamics of three coupled atomic
  Bose-Einstein condensates}},\ }\href
  {https://doi.org/10.1103/PhysRevA.63.013604} {\bibfield  {journal} {\bibinfo
  {journal} {Phys. Rev. A}\ }\textbf {\bibinfo {volume} {63}},\ \bibinfo
  {pages} {013604} (\bibinfo {year} {2000})}\BibitemShut {NoStop}%
\bibitem [{\citenamefont {Franzosi}\ and\ \citenamefont
  {Penna}(2003)}]{Franzosi2003}%
  \BibitemOpen
  \bibfield  {author} {\bibinfo {author} {\bibfnamefont {R.}~\bibnamefont
  {Franzosi}}\ and\ \bibinfo {author} {\bibfnamefont {V.}~\bibnamefont
  {Penna}},\ }\bibfield  {title} {\bibinfo {title} {{Chaotic behavior,
  collective modes, and self-trapping in the dynamics of three coupled
  Bose-Einstein condensates}},\ }\href
  {https://doi.org/10.1103/PhysRevE.67.046227} {\bibfield  {journal} {\bibinfo
  {journal} {Phys. Rev. E}\ }\textbf {\bibinfo {volume} {67}},\ \bibinfo
  {pages} {046227} (\bibinfo {year} {2003})}\BibitemShut {NoStop}%
\bibitem [{\citenamefont {Mossmann}\ and\ \citenamefont
  {Jung}(2006)}]{Mossmann2006}%
  \BibitemOpen
  \bibfield  {author} {\bibinfo {author} {\bibfnamefont {S.}~\bibnamefont
  {Mossmann}}\ and\ \bibinfo {author} {\bibfnamefont {C.}~\bibnamefont
  {Jung}},\ }\bibfield  {title} {\bibinfo {title} {Semiclassical approach to
  {Bose-Einstein} condensates in a triple well potential},\ }\href
  {https://doi.org/10.1103/PhysRevA.74.033601} {\bibfield  {journal} {\bibinfo
  {journal} {Phys. Rev. A}\ }\textbf {\bibinfo {volume} {74}},\ \bibinfo
  {pages} {033601} (\bibinfo {year} {2006})}\BibitemShut {NoStop}%
\bibitem [{\citenamefont {Oelkers}\ and\ \citenamefont
  {Links}(2007)}]{Oelkers2007}%
  \BibitemOpen
  \bibfield  {author} {\bibinfo {author} {\bibfnamefont {N.}~\bibnamefont
  {Oelkers}}\ and\ \bibinfo {author} {\bibfnamefont {J.}~\bibnamefont
  {Links}},\ }\bibfield  {title} {\bibinfo {title} {{Ground-state properties of
  the attractive one-dimensional Bose-Hubbard model}},\ }\href
  {https://doi.org/10.1103/PhysRevB.75.115119} {\bibfield  {journal} {\bibinfo
  {journal} {Phys. Rev. B}\ }\textbf {\bibinfo {volume} {75}},\ \bibinfo
  {pages} {115119} (\bibinfo {year} {2007})}\BibitemShut {NoStop}%
\bibitem [{\citenamefont {Kanamoto}\ \emph {et~al.}(2005)\citenamefont
  {Kanamoto}, \citenamefont {Saito},\ and\ \citenamefont
  {Ueda}}]{Kanamoto2005}%
  \BibitemOpen
  \bibfield  {author} {\bibinfo {author} {\bibfnamefont {R.}~\bibnamefont
  {Kanamoto}}, \bibinfo {author} {\bibfnamefont {H.}~\bibnamefont {Saito}},\
  and\ \bibinfo {author} {\bibfnamefont {M.}~\bibnamefont {Ueda}},\ }\bibfield
  {title} {\bibinfo {title} {{Symmetry Breaking and Enhanced Condensate
  Fraction in a Matter-Wave Bright Soliton}},\ }\href
  {https://doi.org/10.1103/PhysRevLett.94.090404} {\bibfield  {journal}
  {\bibinfo  {journal} {Phys. Rev. Lett.}\ }\textbf {\bibinfo {volume} {94}},\
  \bibinfo {pages} {090404} (\bibinfo {year} {2005})}\BibitemShut {NoStop}%
\bibitem [{\citenamefont {Khaykovich}\ \emph {et~al.}(2002)\citenamefont
  {Khaykovich}, \citenamefont {Schreck}, \citenamefont {Ferrari}, \citenamefont
  {Bourdel}, \citenamefont {Cubizolles}, \citenamefont {Carr}, \citenamefont
  {Castin},\ and\ \citenamefont {Salomon}}]{Khaykovich2002}%
  \BibitemOpen
  \bibfield  {author} {\bibinfo {author} {\bibfnamefont {L.}~\bibnamefont
  {Khaykovich}}, \bibinfo {author} {\bibfnamefont {F.}~\bibnamefont {Schreck}},
  \bibinfo {author} {\bibfnamefont {G.}~\bibnamefont {Ferrari}}, \bibinfo
  {author} {\bibfnamefont {T.}~\bibnamefont {Bourdel}}, \bibinfo {author}
  {\bibfnamefont {J.}~\bibnamefont {Cubizolles}}, \bibinfo {author}
  {\bibfnamefont {L.~D.}\ \bibnamefont {Carr}}, \bibinfo {author}
  {\bibfnamefont {Y.}~\bibnamefont {Castin}},\ and\ \bibinfo {author}
  {\bibfnamefont {C.}~\bibnamefont {Salomon}},\ }\bibfield  {title} {\bibinfo
  {title} {{Formation of a matter-wave bright soliton}},\ }\href
  {https://doi.org/10.1126/science.1071021} {\bibfield  {journal} {\bibinfo
  {journal} {Science}\ }\textbf {\bibinfo {volume} {296}},\ \bibinfo {pages}
  {1290} (\bibinfo {year} {2002})}\BibitemShut {NoStop}%
\bibitem [{\citenamefont {Akila}\ \emph {et~al.}(2017)\citenamefont {Akila},
  \citenamefont {Waltner}, \citenamefont {Gutkin}, \citenamefont {Braun},\ and\
  \citenamefont {Guhr}}]{Akila2017}%
  \BibitemOpen
  \bibfield  {author} {\bibinfo {author} {\bibfnamefont {M.}~\bibnamefont
  {Akila}}, \bibinfo {author} {\bibfnamefont {D.}~\bibnamefont {Waltner}},
  \bibinfo {author} {\bibfnamefont {B.}~\bibnamefont {Gutkin}}, \bibinfo
  {author} {\bibfnamefont {P.}~\bibnamefont {Braun}},\ and\ \bibinfo {author}
  {\bibfnamefont {T.}~\bibnamefont {Guhr}},\ }\bibfield  {title} {\bibinfo
  {title} {{Semiclassical Identification of Periodic Orbits in a Quantum
  Many-Body System}},\ }\href {https://doi.org/10.1103/PhysRevLett.118.164101}
  {\bibfield  {journal} {\bibinfo  {journal} {Phys. Rev. Lett.}\ }\textbf
  {\bibinfo {volume} {118}},\ \bibinfo {pages} {164101} (\bibinfo {year}
  {2017})}\BibitemShut {NoStop}%
\bibitem [{\citenamefont {Franson}\ and\ \citenamefont
  {Donegan}(2002)}]{Franson2002}%
  \BibitemOpen
  \bibfield  {author} {\bibinfo {author} {\bibfnamefont {J.~D.}\ \bibnamefont
  {Franson}}\ and\ \bibinfo {author} {\bibfnamefont {M.~M.}\ \bibnamefont
  {Donegan}},\ }\bibfield  {title} {\bibinfo {title} {Perturbation theory for
  quantum-mechanical observables},\ }\href
  {https://doi.org/10.1103/PhysRevA.65.052107} {\bibfield  {journal} {\bibinfo
  {journal} {Phys. Rev. A}\ }\textbf {\bibinfo {volume} {65}},\ \bibinfo
  {pages} {052107} (\bibinfo {year} {2002})}\BibitemShut {NoStop}%
\bibitem [{Note2()}]{Note2}%
  \BibitemOpen
  \bibinfo {note} {If $\protect \hat {A}$ is a power series in $\hbar
  _{\protect \mathrm {eff}}$ the result applies to the $\hbar _{\protect
  \mathrm {eff}}$-independent coefficients individually.}\BibitemShut {Stop}%
\bibitem [{\citenamefont {Larkin}\ and\ \citenamefont
  {Ovchinnikov}(1969)}]{Larkin1969}%
  \BibitemOpen
  \bibfield  {author} {\bibinfo {author} {\bibfnamefont {A.~I.}\ \bibnamefont
  {Larkin}}\ and\ \bibinfo {author} {\bibfnamefont {Y.~N.}\ \bibnamefont
  {Ovchinnikov}},\ }\bibfield  {title} {\bibinfo {title} {{Quasiclassical
  method in the theory of superconductivity}},\ }\href
  {http://www.jetp.ac.ru/cgi-bin/e/index/e/28/6/p1200?a=list} {\bibfield
  {journal} {\bibinfo  {journal} {Sov. Phys. JETP}\ }\textbf {\bibinfo {volume}
  {28}},\ \bibinfo {pages} {1200} (\bibinfo {year} {1969})}\BibitemShut
  {NoStop}%
\bibitem [{\citenamefont {Rozenbaum}\ \emph {et~al.}(2017)\citenamefont
  {Rozenbaum}, \citenamefont {Ganeshan},\ and\ \citenamefont
  {Galitski}}]{Rozenbaum2017}%
  \BibitemOpen
  \bibfield  {author} {\bibinfo {author} {\bibfnamefont {E.~B.}\ \bibnamefont
  {Rozenbaum}}, \bibinfo {author} {\bibfnamefont {S.}~\bibnamefont
  {Ganeshan}},\ and\ \bibinfo {author} {\bibfnamefont {V.}~\bibnamefont
  {Galitski}},\ }\bibfield  {title} {\bibinfo {title} {{Lyapunov Exponent and
  Out-of-Time-Ordered Correlator's Growth Rate in a Chaotic System}},\ }\href
  {https://doi.org/10.1103/PhysRevLett.118.086801} {\bibfield  {journal}
  {\bibinfo  {journal} {Phys. Rev. Lett.}\ }\textbf {\bibinfo {volume} {118}},\
  \bibinfo {pages} {086801} (\bibinfo {year} {2017})}\BibitemShut {NoStop}%
\bibitem [{\citenamefont {Cotler}\ \emph {et~al.}(2018)\citenamefont {Cotler},
  \citenamefont {Ding},\ and\ \citenamefont {Penington}}]{Cotler2018}%
  \BibitemOpen
  \bibfield  {author} {\bibinfo {author} {\bibfnamefont {J.~S.}\ \bibnamefont
  {Cotler}}, \bibinfo {author} {\bibfnamefont {D.}~\bibnamefont {Ding}},\ and\
  \bibinfo {author} {\bibfnamefont {G.~R.}\ \bibnamefont {Penington}},\
  }\bibfield  {title} {\bibinfo {title} {{Out-of-time-order operators and the
  butterfly effect}},\ }\href {https://doi.org/10.1016/j.aop.2018.07.020}
  {\bibfield  {journal} {\bibinfo  {journal} {Ann. Phys. (Amsterdam)}\ }\textbf
  {\bibinfo {volume} {396}},\ \bibinfo {pages} {318} (\bibinfo {year}
  {2018})}\BibitemShut {NoStop}%
\bibitem [{\citenamefont {Lerose}\ and\ \citenamefont
  {Pappalardi}(2020)}]{Lerose2020}%
  \BibitemOpen
  \bibfield  {author} {\bibinfo {author} {\bibfnamefont {A.}~\bibnamefont
  {Lerose}}\ and\ \bibinfo {author} {\bibfnamefont {S.}~\bibnamefont
  {Pappalardi}},\ }\bibfield  {title} {\bibinfo {title} {{Bridging entanglement
  dynamics and chaos in semiclassical systems}},\ }\href
  {https://doi.org/10.1103/PhysRevA.102.032404} {\bibfield  {journal} {\bibinfo
   {journal} {Phys. Rev. A}\ }\textbf {\bibinfo {volume} {102}},\ \bibinfo
  {pages} {032404} (\bibinfo {year} {2020})}\BibitemShut {NoStop}%
\bibitem [{\citenamefont {Gu}\ and\ \citenamefont {Kitaev}(2019)}]{Gu2019}%
  \BibitemOpen
  \bibfield  {author} {\bibinfo {author} {\bibfnamefont {Y.}~\bibnamefont
  {Gu}}\ and\ \bibinfo {author} {\bibfnamefont {A.}~\bibnamefont {Kitaev}},\
  }\bibfield  {title} {\bibinfo {title} {{On the relation between the magnitude
  and exponent of OTOCs}},\ }\href {https://doi.org/10.1007/JHEP02(2019)075}
  {\bibfield  {journal} {\bibinfo  {journal} {J. High Energy Phys.}\ }\textbf
  {\bibinfo {volume} {2019}}\bibinfo  {number} { (2)},\ \bibinfo {pages}
  {75}}\BibitemShut {NoStop}%
\bibitem [{\citenamefont {Kobrin}\ \emph {et~al.}(2021)\citenamefont {Kobrin},
  \citenamefont {Yang}, \citenamefont {Kahanamoku-Meyer}, \citenamefont
  {Olund}, \citenamefont {Moore}, \citenamefont {Stanford},\ and\ \citenamefont
  {Yao}}]{Kobrin2020}%
  \BibitemOpen
\bibfield  {number} {  }\bibfield  {author} {\bibinfo {author} {\bibfnamefont
  {B.}~\bibnamefont {Kobrin}}, \bibinfo {author} {\bibfnamefont
  {Z.}~\bibnamefont {Yang}}, \bibinfo {author} {\bibfnamefont {G.~D.}\
  \bibnamefont {Kahanamoku-Meyer}}, \bibinfo {author} {\bibfnamefont {C.~T.}\
  \bibnamefont {Olund}}, \bibinfo {author} {\bibfnamefont {J.~E.}\ \bibnamefont
  {Moore}}, \bibinfo {author} {\bibfnamefont {D.}~\bibnamefont {Stanford}},\
  and\ \bibinfo {author} {\bibfnamefont {N.~Y.}\ \bibnamefont {Yao}},\
  }\bibfield  {title} {\bibinfo {title} {{Many-Body Chaos in the
  Sachdev-Ye-Kitaev Model}},\ }\bibfield  {journal} {\bibinfo  {journal} {Phys.
  Rev. Lett.}\ }\textbf {\bibinfo {volume} {126}},\ \href
  {https://doi.org/10.1103/PhysRevLett.126.030602}
  {10.1103/PhysRevLett.126.030602} (\bibinfo {year} {2021})\BibitemShut
  {NoStop}%
\bibitem [{\citenamefont {Yao}\ \emph {et~al.}()\citenamefont {Yao},
  \citenamefont {Grusdt}, \citenamefont {Swingle}, \citenamefont {Lukin},
  \citenamefont {Stamper-Kurn}, \citenamefont {Moore},\ and\ \citenamefont
  {Demler}}]{Yao2016}%
  \BibitemOpen
  \bibfield  {author} {\bibinfo {author} {\bibfnamefont {N.~Y.}\ \bibnamefont
  {Yao}}, \bibinfo {author} {\bibfnamefont {F.}~\bibnamefont {Grusdt}},
  \bibinfo {author} {\bibfnamefont {B.}~\bibnamefont {Swingle}}, \bibinfo
  {author} {\bibfnamefont {M.~D.}\ \bibnamefont {Lukin}}, \bibinfo {author}
  {\bibfnamefont {D.~M.}\ \bibnamefont {Stamper-Kurn}}, \bibinfo {author}
  {\bibfnamefont {J.~E.}\ \bibnamefont {Moore}},\ and\ \bibinfo {author}
  {\bibfnamefont {E.~A.}\ \bibnamefont {Demler}},\ }\href
  {http://arxiv.org/abs/1607.01801} {\bibinfo {title} {{Interferometric
  approach to probing fast scrambling}}},\ \Eprint
  {https://arxiv.org/abs/1607.01801} {arXiv:1607.01801} \BibitemShut {NoStop}%
\bibitem [{\citenamefont {Bohrdt}\ \emph {et~al.}(2017)\citenamefont {Bohrdt},
  \citenamefont {Mendl}, \citenamefont {Endres},\ and\ \citenamefont
  {Knap}}]{Bohrdt2017}%
  \BibitemOpen
  \bibfield  {author} {\bibinfo {author} {\bibfnamefont {A.}~\bibnamefont
  {Bohrdt}}, \bibinfo {author} {\bibfnamefont {C.~B.}\ \bibnamefont {Mendl}},
  \bibinfo {author} {\bibfnamefont {M.}~\bibnamefont {Endres}},\ and\ \bibinfo
  {author} {\bibfnamefont {M.}~\bibnamefont {Knap}},\ }\bibfield  {title}
  {\bibinfo {title} {{Scrambling and thermalization in a diffusive quantum
  many-body system}},\ }\href {http://stacks.iop.org/1367-2630/19/i=6/a=063001}
  {\bibfield  {journal} {\bibinfo  {journal} {New J. Phys.}\ }\textbf {\bibinfo
  {volume} {19}},\ \bibinfo {pages} {063001} (\bibinfo {year}
  {2017})}\BibitemShut {NoStop}%
\bibitem [{\citenamefont {Shen}\ \emph {et~al.}(2017)\citenamefont {Shen},
  \citenamefont {Zhang}, \citenamefont {Fan},\ and\ \citenamefont
  {Zhai}}]{Shen2017}%
  \BibitemOpen
  \bibfield  {author} {\bibinfo {author} {\bibfnamefont {H.}~\bibnamefont
  {Shen}}, \bibinfo {author} {\bibfnamefont {P.}~\bibnamefont {Zhang}},
  \bibinfo {author} {\bibfnamefont {R.}~\bibnamefont {Fan}},\ and\ \bibinfo
  {author} {\bibfnamefont {H.}~\bibnamefont {Zhai}},\ }\bibfield  {title}
  {\bibinfo {title} {{Out-of-time-order correlation at a quantum phase
  transition}},\ }\href {https://doi.org/10.1103/PhysRevB.96.054503} {\bibfield
   {journal} {\bibinfo  {journal} {Phys. Rev. B}\ }\textbf {\bibinfo {volume}
  {96}},\ \bibinfo {pages} {054503} (\bibinfo {year} {2017})}\BibitemShut
  {NoStop}%
\bibitem [{\citenamefont {Keleş}\ \emph {et~al.}(2019)\citenamefont {Keleş},
  \citenamefont {Zhao},\ and\ \citenamefont {Liu}}]{Keles2019}%
  \BibitemOpen
  \bibfield  {author} {\bibinfo {author} {\bibfnamefont {A.}~\bibnamefont
  {Keleş}}, \bibinfo {author} {\bibfnamefont {E.}~\bibnamefont {Zhao}},\ and\
  \bibinfo {author} {\bibfnamefont {W.~V.}\ \bibnamefont {Liu}},\ }\bibfield
  {title} {\bibinfo {title} {{Scrambling dynamics and many-body chaos in a
  random dipolar spin model}},\ }\href
  {https://doi.org/10.1103/PhysRevA.99.053620} {\bibfield  {journal} {\bibinfo
  {journal} {Phys. Rev. A}\ }\textbf {\bibinfo {volume} {99}},\ \bibinfo
  {pages} {053620} (\bibinfo {year} {2019})}\BibitemShut {NoStop}%
\bibitem [{\citenamefont {Lantagne-Hurtubise}\ \emph
  {et~al.}(2020)\citenamefont {Lantagne-Hurtubise}, \citenamefont {Plugge},
  \citenamefont {Can},\ and\ \citenamefont {Franz}}]{Lantagne-Hurtubise2020}%
  \BibitemOpen
  \bibfield  {author} {\bibinfo {author} {\bibfnamefont {{\'{E}}.}~\bibnamefont
  {Lantagne-Hurtubise}}, \bibinfo {author} {\bibfnamefont {S.}~\bibnamefont
  {Plugge}}, \bibinfo {author} {\bibfnamefont {O.}~\bibnamefont {Can}},\ and\
  \bibinfo {author} {\bibfnamefont {M.}~\bibnamefont {Franz}},\ }\bibfield
  {title} {\bibinfo {title} {{Diagnosing quantum chaos in many-body systems
  using entanglement as a resource}},\ }\href
  {https://doi.org/10.1103/PhysRevResearch.2.013254} {\bibfield  {journal}
  {\bibinfo  {journal} {Phys. Rev. Research}\ }\textbf {\bibinfo {volume}
  {2}},\ \bibinfo {pages} {013254} (\bibinfo {year} {2020})}\BibitemShut
  {NoStop}%
\bibitem [{\citenamefont {Heller}(1984)}]{Heller1984}%
  \BibitemOpen
  \bibfield  {author} {\bibinfo {author} {\bibfnamefont {E.~J.}\ \bibnamefont
  {Heller}},\ }\bibfield  {title} {\bibinfo {title} {{Bound-State
  Eigenfunctions of Classically Chaotic Hamiltonian Systems: Scars of Periodic
  Orbits}},\ }\href {https://doi.org/10.1103/PhysRevLett.53.1515} {\bibfield
  {journal} {\bibinfo  {journal} {Phys. Rev. Lett.}\ }\textbf {\bibinfo
  {volume} {53}},\ \bibinfo {pages} {1515} (\bibinfo {year}
  {1984})}\BibitemShut {NoStop}%
\end{thebibliography}%
\end{document}